\documentclass[amsmath,amssymb,superscriptaddress,twocolumn]{revtex4-2}
\usepackage[utf8]{inputenc}
\usepackage{hyperref}
\usepackage{graphicx}
\usepackage{dcolumn}
\usepackage{bm}
\usepackage{xcolor}
\usepackage[normalem]{ulem}

\begin{document}

\title{Emergence of Disordered Hyperuniformity in Confined Fluids and Soft Matter}

\author{Fabio Leoni}
\email{fabio.leoni@uniroma1.it}
\affiliation{Dipartimento di Fisica, Universit\`a degli Studi di Roma La Sapienza, Piazzale Aldo Moro 5, Rome, 00185, Italy}

\author{Erdal C. O\u{g}uz}
\email{ecoguz@iphy.ac.cn} 
\affiliation{Key Laboratory of Soft Matter Physics, Institute of Physics, Chinese Academy of Sciences, Beijing, 100190, China}
\affiliation{Songshan Lake Materials Laboratory, Dongguan, Guangdong 523808, China}
\affiliation{School of Chemistry and the Center for Physics and Chemistry of Living Systems, Tel Aviv University, Tel Aviv 6997801, Israel}

\author{Giancarlo Franzese}
\email{gfranzese@ub.edu}
\affiliation{Secci\'o de F\'isica Estad\'istica i Interdisciplin\`aria - Departament de F\'isica de la Mat\`eria Condensada, Universitat de
  Barcelona, Mart\'{\i} i Franqu\`es 1, 08028 Barcelona, Spain.\\
Institut de Nanoci\`{e}ncia i Nanotecnologia, Universitat de Barcelona, Barcelona, Spain.}

\date{\today}

\begin{abstract}
Confinement can significantly alter fluid properties, offering potential for specific technological applications. However, achieving precise control over the structural complexity of confined fluids and soft matter remains challenging, as it often requires careful tuning of system parameters.
In this study, we perform large-scale molecular dynamics simulations of a fluid confined in a slit pore, employing an interaction potential applicable to a broad range of soft matter. Confinement induces the fluid to form two-dimensional (2D) layers, each of which self-assembles into structures of varying complexity. Notably, we observe a suppression of large-scale density fluctuations—akin to crystals—within complex nontrivial fluid-like geometries, leading to the emergence of disordered hyperuniformity.
We demonstrate that fluid properties can be precisely adjusted by modifying control parameters. This work provides a foundation for designing experimental protocols to regulate the structure of self-assembling materials, stabilize disordered hyperuniform structures, and facilitate the fabrication of 2D arrays with tailored complexity. Additionally, it offers a new perspective for analyzing biological monolayers, such as epithelial tissues.
\end{abstract}

\maketitle

%
%

The properties of liquids and colloids change significantly under confinement, displaying altered thermodynamic, dynamic, and structural characteristics due to the coupling with the boundary~\cite{drake90,leoni2014,leoni2021,mittal2008,stewart2012,ian15,zaragoza2019molecular,krott2013}. 
In these confined systems, the nature of the confinement exerts substantial control over liquid behavior. For instance, fluids near attractive walls exhibit increased density, leading to an increased freezing temperature relative to bulk conditions. In contrast, repulsive walls increase local density fluctuations, thereby lowering the freezing point \cite{alba06}. 

Confinement geometry also shapes the ordering of mesoscopic systems near boundaries. Colloids within cylindrical, circular, and spherical enclosures self-assemble into helical, concentric, or icosahedral configurations, respectively \cite{mughal12,ian13,nijs15,chen21}. 
In slit-like geometries with parallel walls, particles often form layered structures, resulting in oscillatory density profiles perpendicular to the walls \cite{evans1988,lutsko2008}. 
While layering frequently promotes crystallization, structured fluid layers have also been observed, e.g., in water ~\cite{leoni2021,martelli_confined,chiricotto2021role,weiwettability2022,gao2018phase,zaragoza2019molecular}. Soft confinements, such as those in biological systems, generate long-range perturbations in bonding networks that propagate at bulk-like densities ~\cite{martelli_confined}. By contrast, hard confinements like those created by graphene alter contact angles and wettability~\cite{chiricotto2021role,weiwettability2022}.

Unlike crystalline structures, liquids lack conventional long-range order due to the absence of discrete symmetry, commonly resulting in enhanced long-wavelength density fluctuations compared to ordered phases. In crystals, periodic order suppresses density variations over large length scales, causing the static structure factor,  $S(\mathbf{k})$ , to decay to zero at long wavelengths, i.e., $\lim_{|\mathbf{k}| \to 0} S(\mathbf{k}) = 0$ (with $\mathbf{k}$ as the wave vector). Hence, crystals are {\it hyperuniform} \cite{torquato2003local}, a phenomenon sometimes termed superhomogeneous in cosmology \cite{gabrielli2002}.

Conversely, liquids and other disordered states are typically non-hyperuniform, as their $S({\bf k})$  converges to a finite, often substantial value as $k\equiv |\mathbf{k}|\to0$, which correlates with the isothermal compressibility, $\kappa_T$, in equilibrium. The compressibility can be expressed as 
$\kappa_T=(\rho k_BT)^{-1}\lim_{|\mathbf{k}| \to 0} S(\mathbf{k})$, where $\rho$ is density,  $T$  the absolute temperature, and $k_B$ the Boltzmann constant. This ties the concept of hyperuniformity directly to material thermodynamic properties.

Recent studies have identified disordered hyperuniform states—systems that, although statistically isotropic akin to liquids, exhibit a partial reduction in large-scale density fluctuations, resembling crystalline structures. 
Examples include, but are not limited to, sheared emulsions \cite{weijs15}, jammed hard-particle packings \cite{zach11}, critical absorbing states \cite{hexner15}, random organization models \cite{laurent08}, amorphous ices~\cite{martelli2017}, amorphous silicon~\cite{xie2013hyperuniformity},
two-dimensional (2D) amorphous silica~\cite{zheng2020disordered}, active colloidal particles \cite{lei2019}, algae confined to water-air interface \cite{huang21}. 
This disordered hyperuniformity extends our understanding of structural order and its roles in nature. Materials endowed with this property show unique mechanical and optical characteristics~\cite{torquato2018,xu17,gorsky2019engineered,muller2014silicon,milovsevic2019hyperuniform}, suggesting significant potential for future technological applications. Therefore, identifying control parameters to induce hyperuniformity in new materials is an important direction for ongoing research.

In this work, we simulate a model fluid confined within a slit pore and demonstrate that, by adjusting control parameters, self-assembled layered structures of varying complexity can emerge, including hyperuniform, non-hyperuniform, and anti-hyperuniform configurations. The latter refers to long-wavelength density fluctuations exceeding those in random Poissonian media, as indicated by a diverging  $S(k \to 0)$  and commonly observed in systems near critical points \cite{torquato2018}. Our model encompasses a range of materials, including colloids, globular proteins, liquid metals, and, to some extent, water. These findings provide a framework for the experimental realization of (nearly) hyperuniform macroscopic assemblies, paving the way for advanced manufacturing of two-dimensional arrays with diverse structural complexity.

%
%

\vspace{0.5cm}
\noindent{\bf CSW fluid in a slit-pore confinement}

We perform molecular dynamics (MD) simulations of a continuous shouldered well (CSW) fluid confined in a slit pore in the $NVT$ ensemble, where $N$ is the number of particles, $V$ the volume of the simulation box, and $T$ the temperature, by using LAMMPS \cite{LAMMPS}.
\\

\noindent{\bf CSW fluid:} 
The anomalous CSW fluid is described by the effective pair potential between two particles at a distance $r$ \cite{franzese2007, vilaseca2010,leoni2014,leoni2016,deoliveira2008} given as
\begin{equation}\label{eq:eq1}
\begin{array}{lll}
U(r) & \equiv & \dfrac{U_R}{1+\exp(\Delta(r-R_R)/a)}\\
&&\\
&& -U_A\exp\left[-\dfrac{(r-R_A)^2}
  {2\delta_A^2}\right]+U_A\left(\dfrac{a}{r}\right)^{24} ,   
\end{array}
\end{equation}
where $R_A$ and $R_R$ represent the distances of the attractive minimum and the repulsive radius, respectively; $U_A$ and $U_R$ are the energies of the attractive well and the repulsive shoulder, respectively; $\delta_A^2$ is the variance of the Gaussian centered at $R_A$; and $\Delta$ is the parameter controlling the slope between the shoulder and the well at $R_R$.

\begin{figure*}[t!]
\begin{center}
\includegraphics[clip=true,width=11cm]{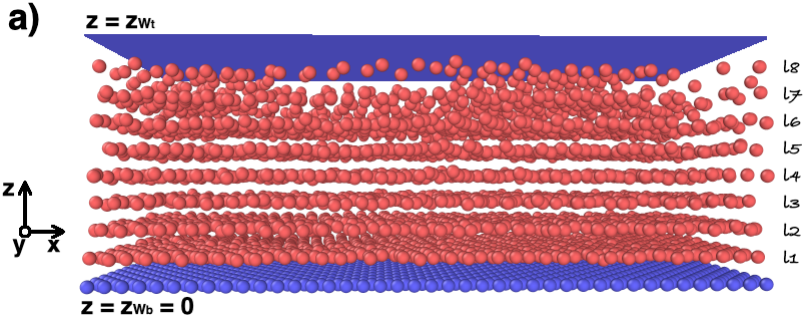}
\hspace{0.5cm}
\includegraphics[clip=true,width=5.7cm]{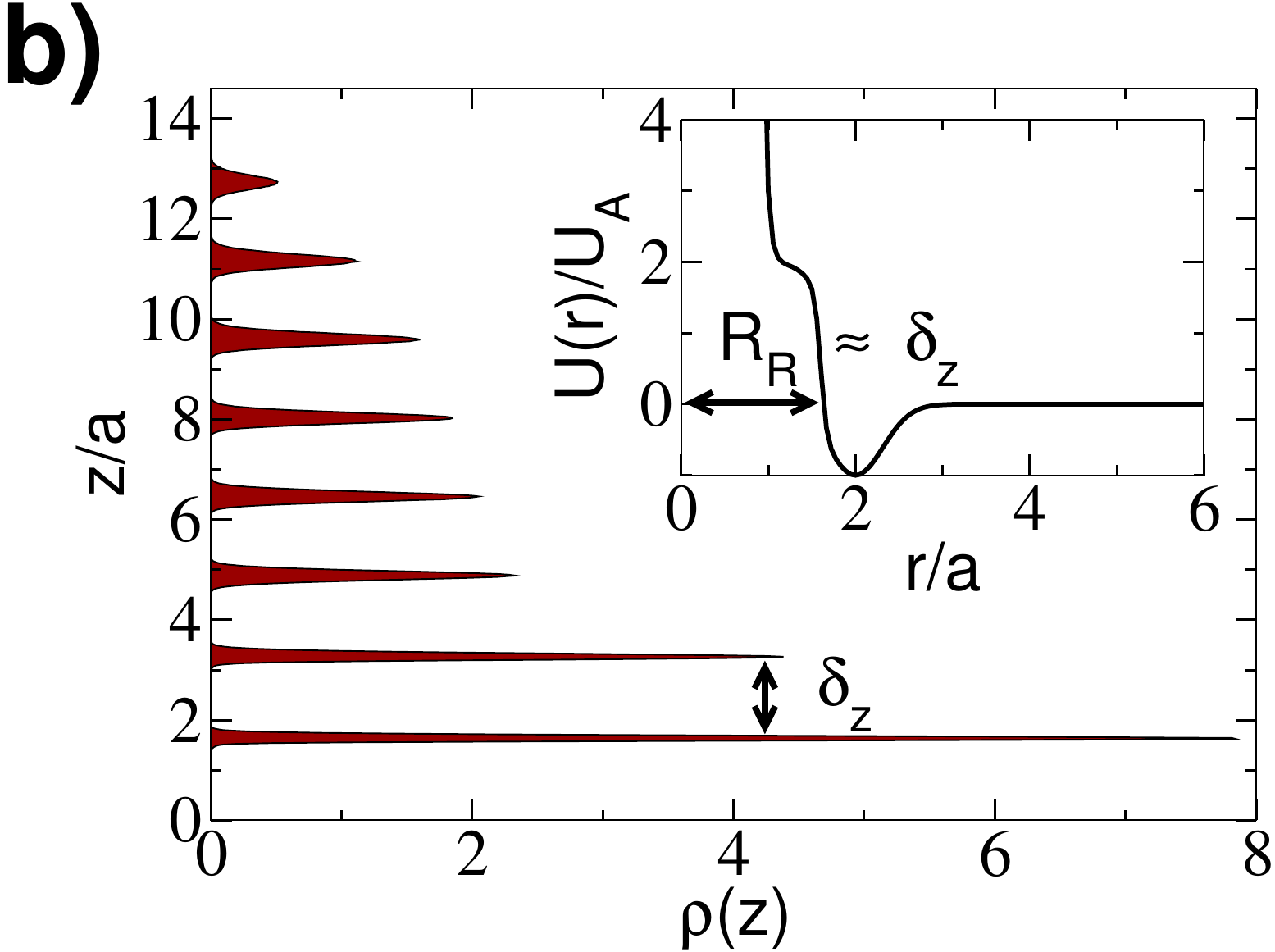}
\vspace{0.5cm}
\includegraphics[clip=true,width=0.7cm]{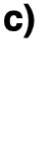}
\includegraphics[clip=true,width=2.05cm]{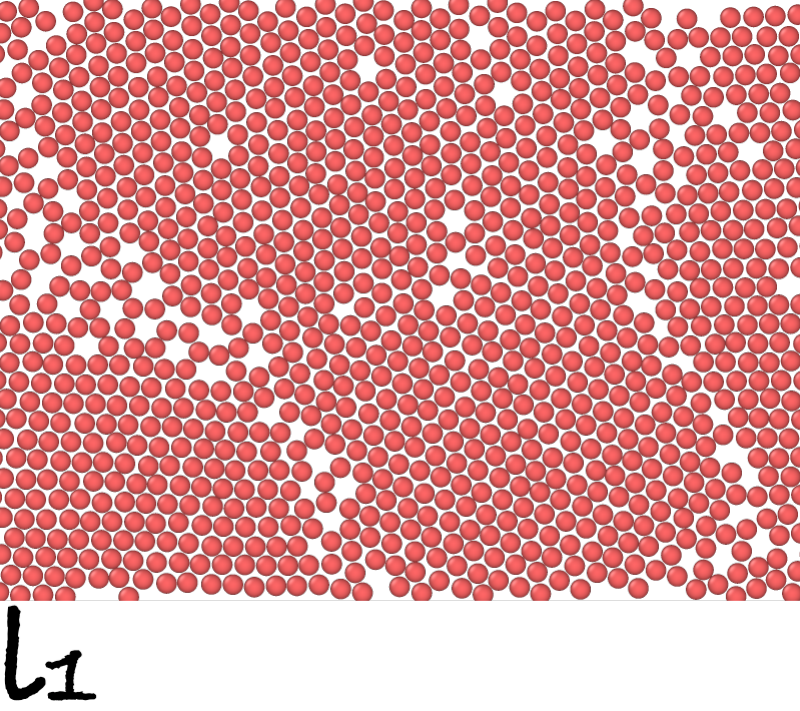}
\includegraphics[clip=true,width=2.05cm]{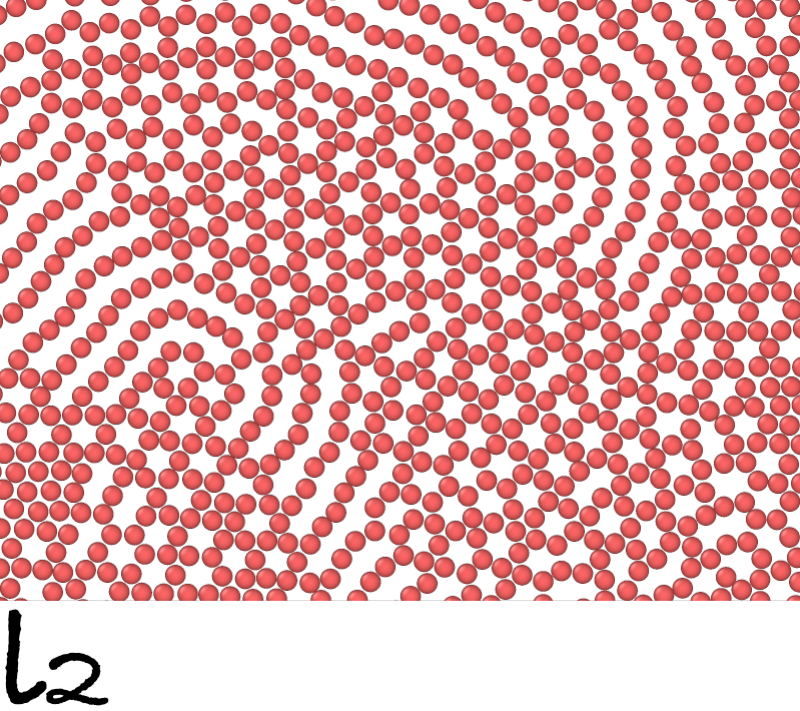}
\includegraphics[clip=true,width=2.05cm]{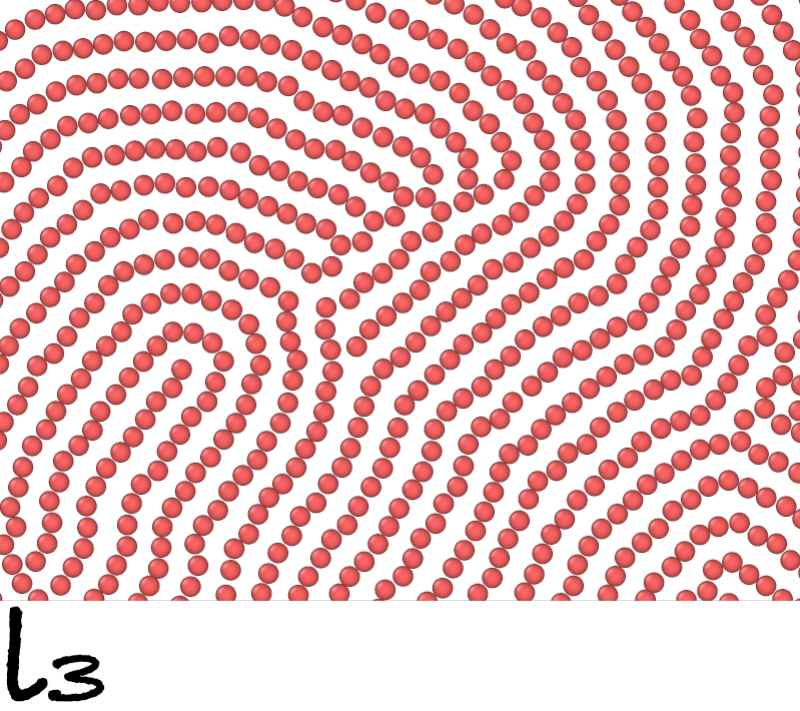}
\includegraphics[clip=true,width=2.05cm]{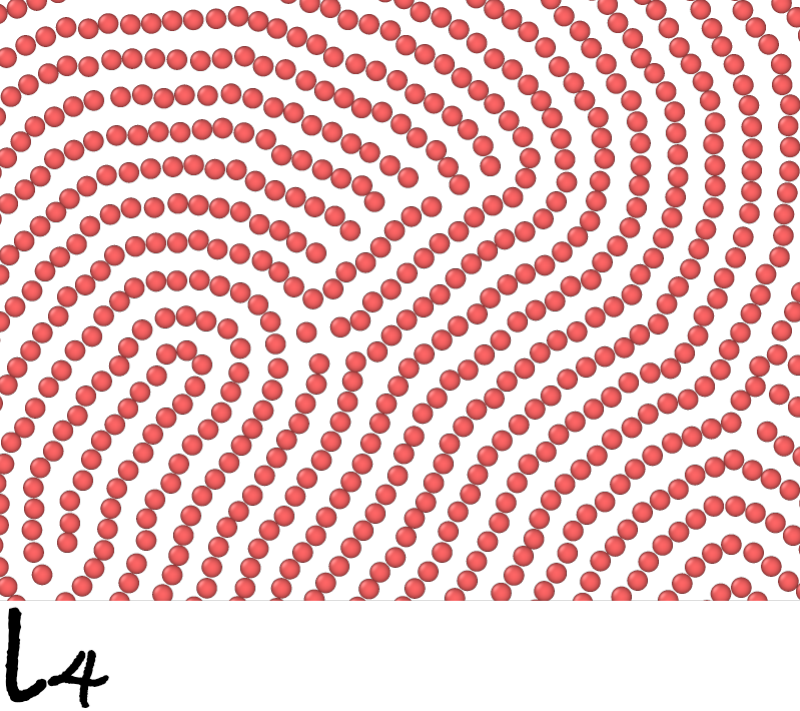}
\includegraphics[clip=true,width=2.05cm]{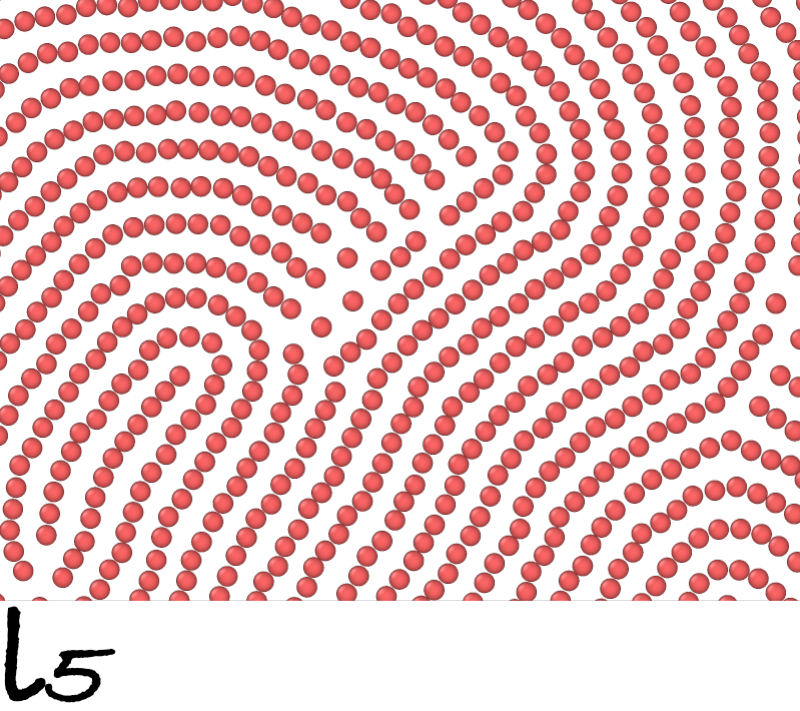}
\includegraphics[clip=true,width=2.05cm]{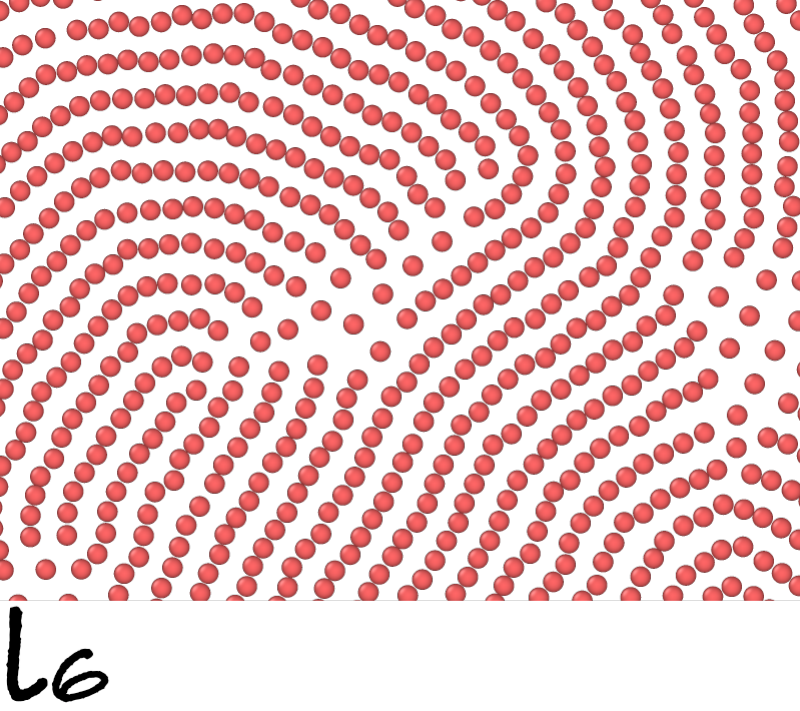}
\includegraphics[clip=true,width=2.05cm]{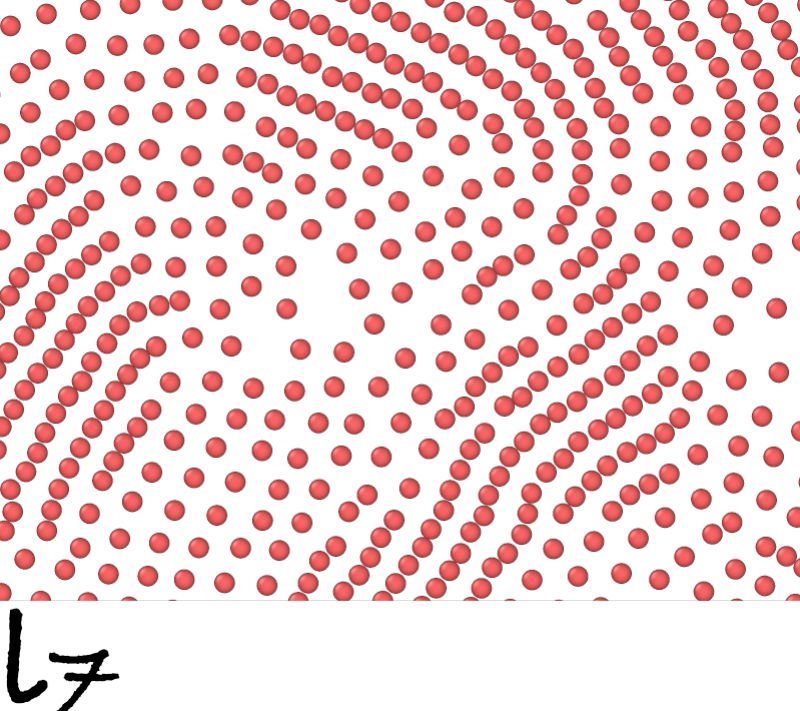}
\includegraphics[clip=true,width=2.05cm]{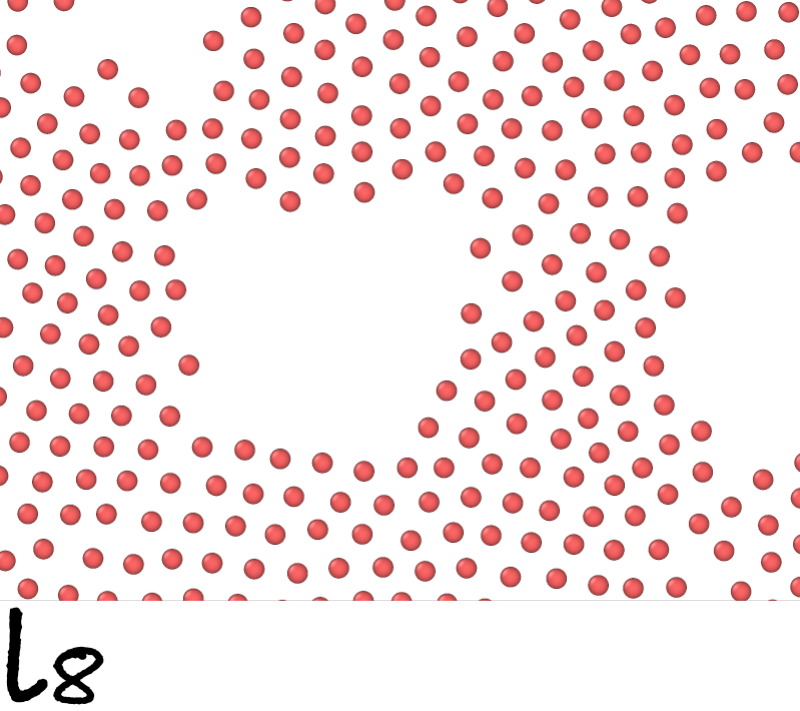}
\end{center}
\caption{{\bf The CSW Fluid under slit-pore confinement}. 
(a) Representative simulation snapshot showing a small section of the confined fluid (red particles) at $\rho = 0.3$  and $T = 0.3$ after rapid annealing and equilibration ($\tau = t_0$). The confinement between two parallel walls (blue: bottom wall attractive, top wall repulsive) produces eight well-defined layers, labeled $l_1, l_2, \dots, l_8$ from the attractive wall outward.
(b) The density profile along the $z$-axis characterizes the layers, with the inter-layer spacing $\delta_z$ approximately matching the repulsive radius  $R_R$ of the CSW potential (inset in panel b). (c) Snapshots of a subregion (50 × 30 in units of $a$) in the  $xy$-plane ($\simeq 2.6\%$ of the total system area) for each layer labeled as $l_i$ (where $i = 1, \dots, 8$), highlighting the particle arrangement within individual layers.
}
\label{fig:fig1}
\end{figure*}

We set the well parameters to be $U_R/U_A=2$, $R_R/a=1.6$, $R_A/a=2.0$, $(\delta_A/a)^2=0.1$, following the approach outlined in Refs.~\cite{vilaseca2010,leoni2014,leoni2016}. Additionally, we adopt $\Delta=30$, consistent with the bulk case studied in Ref.~\cite{vilaseca2010}.
Temperature, density, length, and time are expressed in reduced units: $T^*\equiv K_BT/U_A$, $\rho^*\equiv\rho a^3$, $r^*\equiv r/a$ and $t^*\equiv (a^2m/U_A)^{1/2}$, respectively. The potential profile is shown in the inset of Fig.~\ref{fig:fig1}(b). For clarity, we omit the asterisks denoting reduced units in the following sections unless otherwise specified.
\\

\noindent{\bf Slit pore:} The slit pore consists of two parallel walls in the $xy$-plane, positioned along the $z$-axis at $z_{W_b}=0$ and $z_{W_t}$ for the bottom and top walls, respectively, with a box height along $z$ of $L_z=z_{W_t}-z_{W_b}$.
The bottom wall, modeled as attractive, is formed by CSW particles arranged in a triangular lattice of lattice constant $d=a$, where particle centers are located at $z=0$, following the setup in Refs.~\cite{leoni2014,leoni2016}.
The top wall, modeled as repulsive, is represented by a WCA potential, diverging at $z=L_z$ (i.e., a 12-6 Lennard-Jones potential 
with $\epsilon_W=1$ and $\sigma_W=1$, and a cut-off at the minimum position, $z=L_z-2^{1/6}\sigma_W$, and shifted to zero). 
We assume zero gravity, meaning that the only asymmetry along the  $z$-axis is due to fluid-wall interactions.

The average distance between neighboring fluid layers,  $\delta_z$, is approximately equal to the CSW repulsive diameter $R_R$  (i.e., $\delta_z \approx R_R$). We set $z_{W_t}=(n_l+0.5)\delta_z+\sigma_W$ where $n_l$ represents the maximum number of fluid layers that can form at the highest density $\rho$  for each temperature $T$ considered. Specifically, we analyze three temperatures:  $T = 0.7, 0.5, 0.3$, and two number densities  $\rho = N / V = 0.2, 0.3$, for three systems of size $n_l = 5, 8, 10$  (illustrated, for instance, in Fig.~\ref{fig:fig1}a).
To set the density, we adjust the number of fluid particles $N$ (either 250~000 or 300~000) and the lateral sizes of the simulation box, $L_x$ and $L_y$, with periodic boundary conditions applied along the $x$- and $y$-directions.
Particles belonging to each layer $l_i$ (for $i=1,..,n_l$, where $l_1$ denotes the layer nearest to the bottom wall, $l_2$ the next closest, and so forth, as illustrated in Fig.~\ref{fig:fig1}a) are defined based on their $z$-coordinate. Specifically, particles are assigned to the first layer $l_1$ if $0<z<(3/2)\delta_z$, and to subsequent layers $i>1$ if $(i-1/2)\delta_z<z<(i+1/2)\delta_z$.
\\

\noindent{\bf Simulation protocol:} 
We conduct MD simulations with a time step $dt=0.004$ corresponding apprximantely to $2.6\cdot 10^{-12}$~s for argon-like atoms.
Following the approach in Ref.~\cite{leoni2014,leoni2016}, we start from a liquid phase at a relatively high temperature $T=3$ and implement two different annealing protocols to reach target temperatures of $T=0.7, 0.5$, and $0.3$: a fast annealing procedure with a characteristic time $\tau=t_0$ and a slow annealing with $10t_0$, as outlined below.

In each case, we first cool the system from $T=3$ to $T=1.4$ in $10^5$ timesteps ($t_0=10^5\cdot0.004=400$). We then equilibrate the system at $T=1.4$ for $t=\tau$, followed by an annealing step from  $T=1.4$ to $T=0.7$ over a period of $t=2\tau$.  After reaching $T=0.7$, we equilibrate the system again for  $t=\tau$. This sequence continues, with the system subsequently annealed to  $T=0.5$ over $t=2\tau$, equilibrated at $T=0.5$ for $t=\tau$, further annealed to  $T=0.3$ over $t=2\tau$, and finally equilibrated at $T=0.3$ for $t=\tau$.

%
%

To examine the effect of pore width (wall-to-wall separation distance) on the fluid, we explore three different system sizes, 
allowing the formation of up to $5$, $8$, and $10$ layers at high density and low temperature. 
In the following, we describe the properties of the 8-layer system and refer to other system sizes as needed. 

\vspace{0.5cm}
\newpage
\noindent{\bf Structural diversity in confinement-induced layers}

The 2D number density profile along the wall-to-wall separation coordinate $z$,  $\rho_z(z) = n(z) / (L_x L_y)$, where  $n(z)$ 
is the number of particles at position $z$, shows distinct peaks (Fig.~\ref{fig:fig1}b). With increasing temperature, the peaks broaden and decrease in intensity, eventually forming a uniform profile further from the attractive wall (Fig.~S1).
In the slow annealing protocol ($\tau = 10 t_0$), the first two peaks increase in intensity compared to fast annealing, reflecting a reduction in defects within layers $l_1$ and $l_2$ (Fig.~S2).

By analyzing the snapshots (Fig.~\ref{fig:fig1}c), we observe a templating effect exerted by the attractive wall on the adjacent fluid layer $l_1$, as well as a “molding” effect of $l_1$  on $l_2$, which propagates through subsequent layers ($l_i$ for $i>2$), consistent with findings in Ref.~\cite{leoni2014}. 
The molding effect of layer $l_i$ on layer  $l_{i+1}$ (for  $i \geq 1$) becomes apparent when plotting pairs of adjacent layers (Fig.~S3). For $i\geq2$, each layer occupies the spaces left by the previous layer (with index $i-1$).

At low temperatures, we observe various competing structural arrangements within each layer. For example, at $\rho=0.3$, the fluid in $l_1$ forms a compact triangular lattice (lattice constant $\simeq 1$) containing point-like defects and linear defects (grain boundaries) (Fig.~S4). 
At this same density, layer $l_2$ exhibits a mix of compact triangular lattice, Kagom\'e, honeycomb, and stripe phases (Fig.~S3, third column from the left, second row from the bottom).From $l_3$ to $l_7$, the fluid forms stripes with an increasing presence of low-density triangular lattice regions (lattice constant $\simeq R_R/a$). In layer $l_8$, the only observed structure is a low-density triangular lattice with void defects.
Thus, layers $l_8$ and $l_1$ share isocrystalline structures, displaying the same spatial arrangement but with differing lattice constants. This property arises from the CSW potential’s soft shoulder, which, in bulk, is related to its polymorphic and polyamorphic behaviors. Such features are seen in other isotropic soft-core potentials with two characteristic length scales and anomalous properties \cite{vilaseca2010}.

At $\rho = 0.2$ (Fig.~S3, first column from the left), we observe a similar sequence of structural formations within the layers. However, due to the lower density, the compact triangular lattice in $l_1$ is largely absent, while the low-density triangular lattice begins to emerge in $l_3$.

\vspace{0.5cm}
\noindent{\bf Emergence of disordered hyperuniformity}

\begin{figure}[t!]
\begin{center}
\includegraphics[clip=true,width=8.5cm]{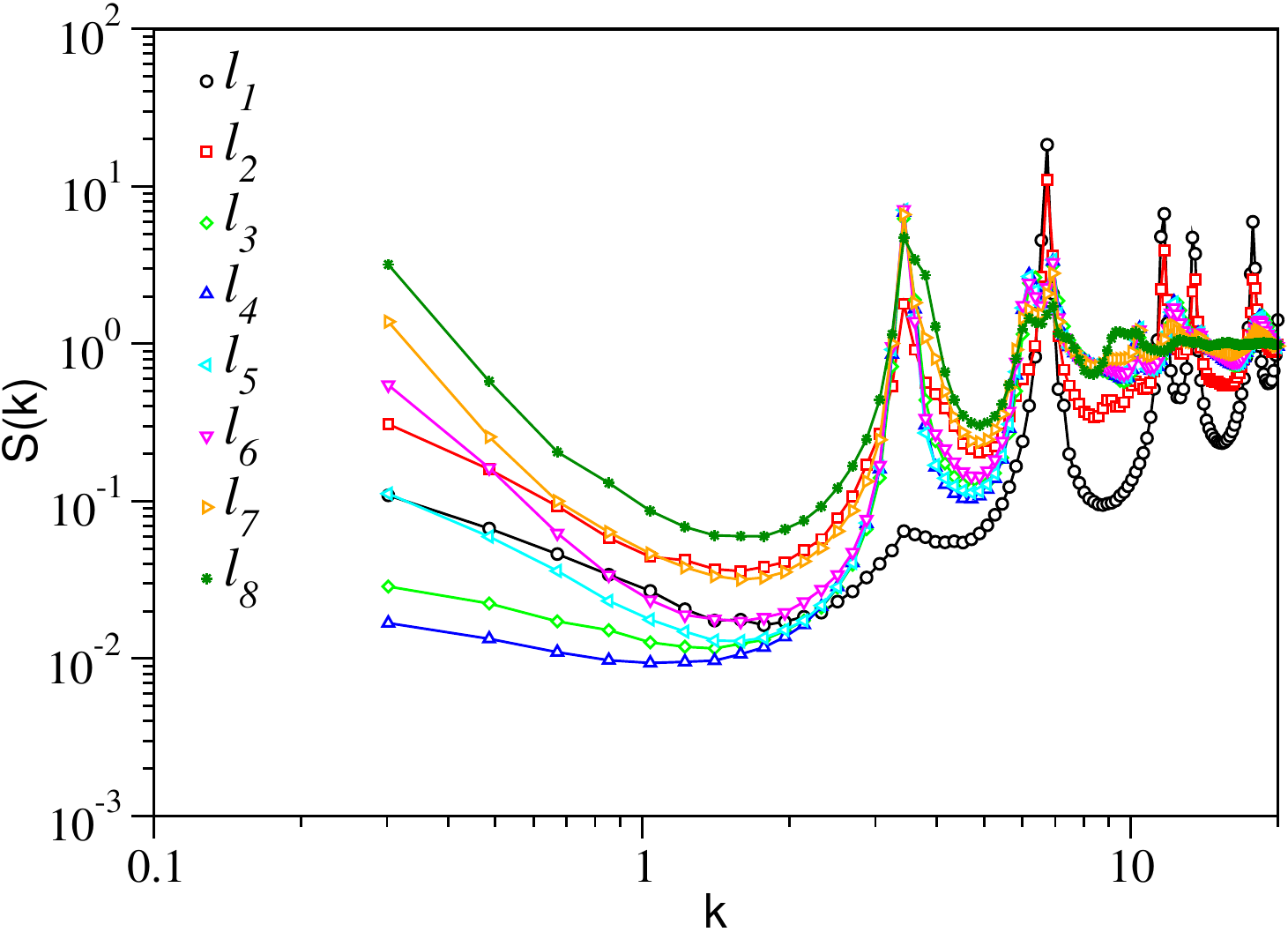}
\end{center}
\caption{{\bf Layer-by-layer structure factor of the confined fluid.} Double-logarithmic plot of the 2D $S(k)$ in the $xy$-plane for each layer $l_i$, from $i=1$ (black open circles) to $8$ (green full circles), with intermediate layers represented by symbols and colors as shown in the legend, for the pore at $\rho=0.3$ and $T=0.3$ in Fig.~\ref{fig:fig1}.}
\label{fig:SF}
\end{figure}

Next, given the distinctive features of each layer and their sharp separation at low $T$ (see Figs.S1, S2), we study the density fluctuations on a layer-by-layer basis. Specifically, for each layer, we calculate the 2D $S(k)$ by projecting the fluid coordinates onto the $xy$-plane (see Fig.~\ref{fig:SF} for the case at $\rho=0.3$, $T=0.3$ and $\tau=t_0$).
We observe that $S(k_p)$, where $k_p$ denotes the wave number associated with the highest structure factor peak, varies monotonically with the layer number. In contrast, $S(k=0)$, which we estimate using a cubic spline interpolation as outlined in Ref.~\cite{hejna2013nearly}, does not follow a monotonic trend at low temperatures ($T=0.3$ and $T=0.5$).
Based on these findings, we calculate the hyperuniformity index $H\equiv S(0)/S(k_p)$ to assess the extent of hyperuniformity within each layer.  
(Systems with $H\lesssim10^{-3}$ are generally considered nearly hyperuniform, with increasing degrees of hyperuniformity at lower values of $H$~\cite{torquato2018,torquato2021}.)

\begin{figure}[t!]
\includegraphics[clip=true,width=4.25cm]{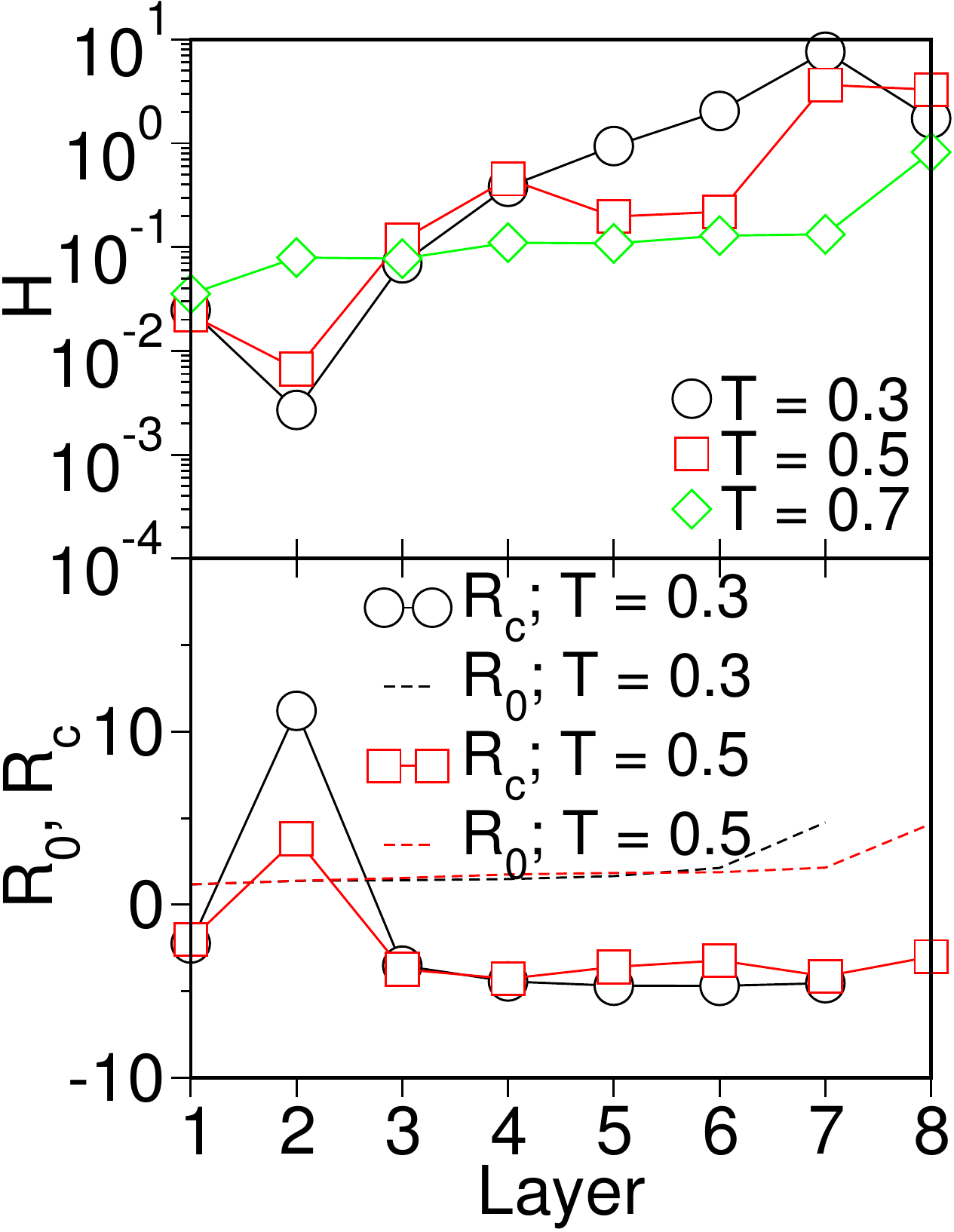} 
\includegraphics[clip=true,width=4.25cm]{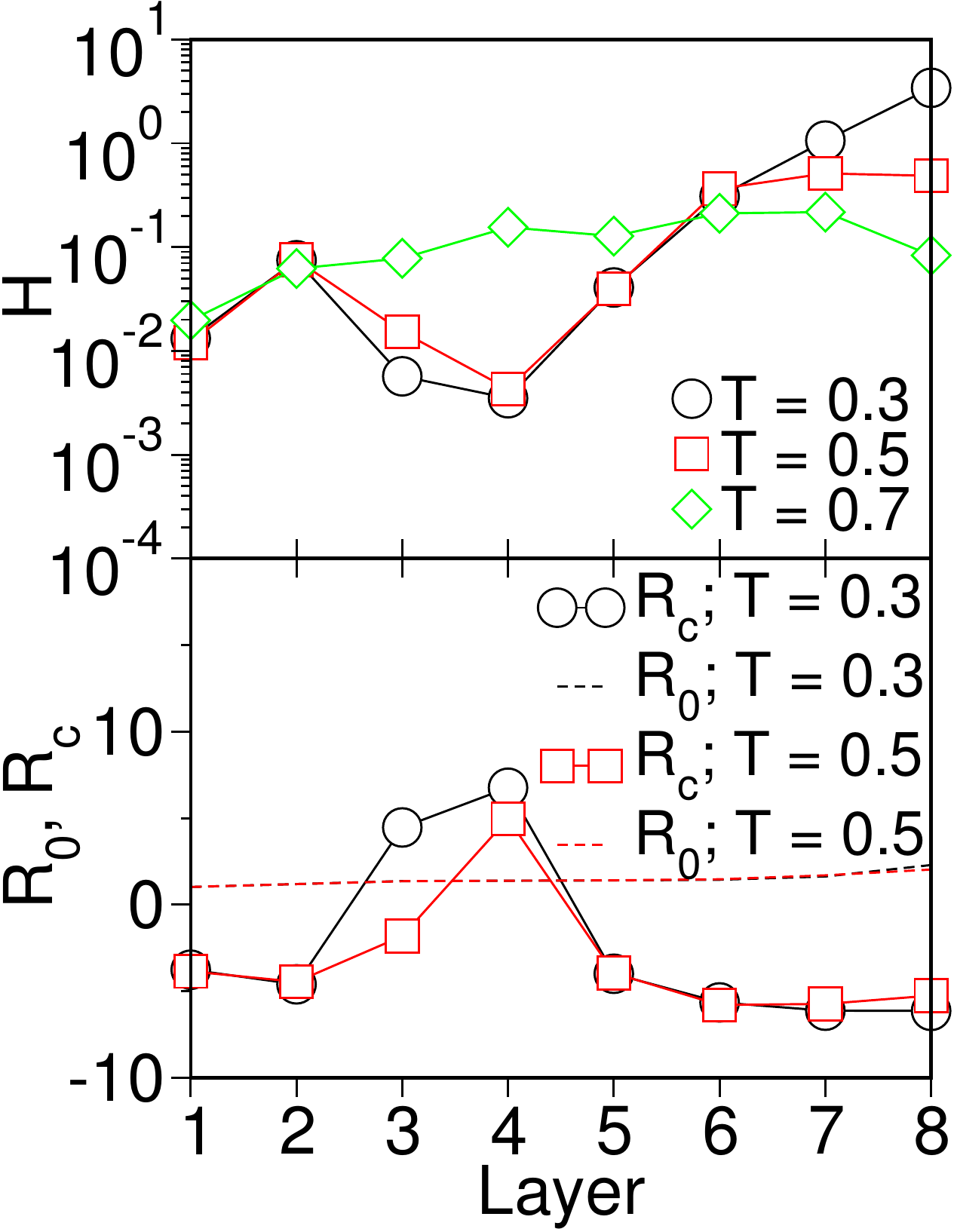} 
\caption{{\bf Layer-by-layer hyperuniformity index $H$ and $R_0$, $R_c$ for slow annealing rate.} Calculations are presented for systems at low ($\rho=0.2$, left panels) and high ($\rho=0.3$, right panels) density. $H$  (upper panels) and $R_0, R_c$ (lower panels) at $T=0.3$ (black circles), $T=0.5$ (red squares), and $T=0.7$ (green rhombuses) only for $H$.}   
\label{fig:H-index}
\end{figure}

As a result, at low $T$, $H(l_i)$ is strongly non-monotonic  (Fig.~\ref{fig:H-index}), varying significantly between $\simeq 10^{-3}$ (nearly-hyperuniform layers) and $\simeq 10$ (nearly anti-hyperuniform layers \cite{torquato2018}).
Visual inspection reveals that layers with $H\lesssim10^{-2}$ typically exhibit either a triangular lattice with defects or a stripe-like geometry (e.g., $l_3$ and $l_4$ in Fig.~\ref{fig:fig1}c). 
Layers with $10^{-2}<H<1$ display phase separation between two structural forms as seen in $l_2$,  $l_5$, and $l_6$ in Fig.~\ref{fig:fig1}c.
In contrast, nearly anti-hyperuniform layers exhibit cavitation (e.g., $l_8$ in Fig.~\ref{fig:fig1}c).

The presence of defects, such as vacancies and dislocations, within some crystalline layers is known to reduce hyperuniformity in proportion to the defect concentration \cite{kim2018}. 
We observe that a $2\%$ concentration of randomly distributed point defects in an otherwise perfect triangular lattice is sufficient to reduce hyperuniformity to levels comparable to those found in certain layers of our system (Fig.~S8). Additionally, correlated defects, such as grain boundaries, further diminish hyperuniformity. At a fixed defect concentration of $2\%$ in a perfect triangular lattice, we find that the $H$-index increases proportionally with the degree of defect correlation (see Figs.~S8, S9).

At higher temperatures ($T=0.7$, Fig.~\ref{fig:H-index}), $H(l_i)$ levels off around $H\simeq 10^{-1}$. In this regime, the thermal noise induces inter-layer diffusion, altering the liquid's layering structure (praticularly for $l_i$ with $i>2$, Fig.S1) and disrupting the nearly hyperuniform character.
We find similar behavior for $H(l_i)$ for systems with a greater number of layers (Fig.S7), but not in those with fewer layers (Fig.S6).

\begin{figure}
\begin{center}
\includegraphics[clip=true,width=8.5cm]{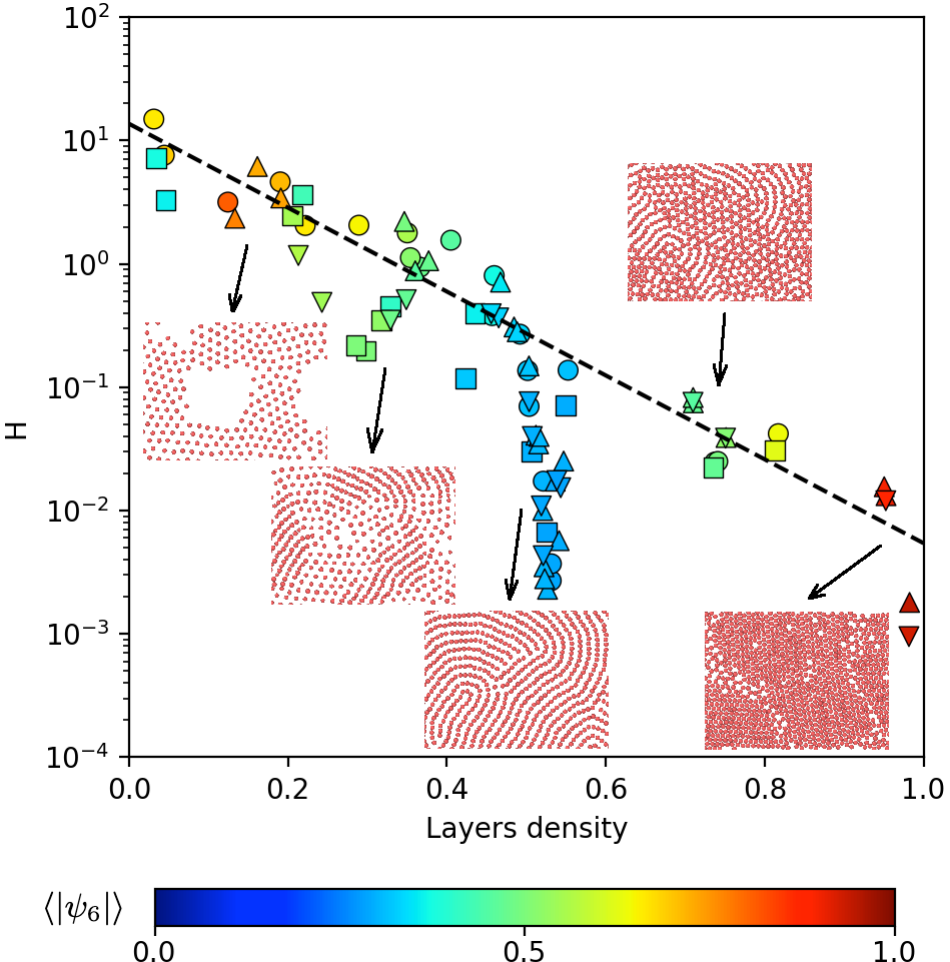}
\end{center}
\caption{{\bf Scaling of the hyperuniformity index $H$ at different layer densities.} Data, color-coded by their $\langle|\psi_6|\rangle$ values, are for systems with $8$ and $10$ layers at $(\rho, T) =(0.2,0.3)$ (circles), $(0.2, 0.5)$ (squares), $(0.3, 0.3)$ (upward triangles), $(0.3, 0.5)$ (downward triangles), and $\tau=t_0$, $10t_0$.  While $H$ is largely uncorrelated with $\langle|\psi_6|\rangle$ within the range $\in [0.4, 1]$, it exhibits an average exponential decrease over five orders of magnitude from a nearly anti-hyperuniform value of $H_0=11.37$ (the zero-density limit for $H$) towards  $H\simeq 10^{-3}$, following $H_{\rm max}=H_0 \exp(-\rho/\rho_1)$ (black dashed line), with $\rho_1 =0.13$. Notably,
at $\rho \simeq 0.5$, $H$ drops to a nearly hyperuniform value when $\langle|\psi_6|\rangle \simeq 0.3$, regardless of $(\rho, T)$. 
The arrows indicate typical snapshots for each thermodynamic condition, showing compact triangular lattices with minimal defects ($\rho=0.3$, $\tau=10t_0$, $\langle|\psi_6|\rangle \simeq 1$) and stripe phases ($\langle|\psi_6|\rangle \simeq 0.3$) associated with emerging hyperuniformity. 
}
\label{fig:H_vs_rho_ALL}
\end{figure}

For many ordered and disordered systems, the asymptotic behavior of the particle number variance at large $R$ can be approximated as the sum of volume and surface terms, as shown in Eq. (S2). Here, $A$ and $B$ are the volume and surface-area coefficients, respectively, associated with each term in Eq. (S2). 
The ratio $B/A$ provides a measure of the crossover distance $R_c\approx B/A$, which distuinguishes between hyperuniform and non-hyperuniform scaling regimes. A system exhibits hyperuniform
scaling within the range $R_0<R<R_c$, where $R_0\equiv \rho^{-1/d}$ represents the average nearest neighbor distance. The lower panels of Fig.\ref{fig:H-index} (and Fig.\ref{fig:H-index_SM}) display the values of $R_0$ and $R_c$ for systems with densities $\rho=0.2$ and $0.3$, under both slow and fast annealing conditions across different temperatures.
Notably, configurations with $H < 10^{-2}$ often show hyperuniform scaling within the range $R_0<R<R_c$, thereby affirming the nearly hyperuniform nature of these samples.

Next, we calculate the sixfold bond-orientational order parameter for each particle $j$, defined as $\psi_6(j)\equiv \frac{1}{N_j}\sum_{h=1}^{N_j}\exp(i6\theta_{jh})$, where the sum is over the first neighbors $h=1, \dots, N_j$ of the particle $j$. Here, $\theta_{jh}$ represents the angle between the 2D vector connecting the particles $j$ and $h$, ${\bf r}_{xy}^{jh}$, and a fixed reference direction in the $xy$-plane. 
When averaged over all layer particles in a layer, $\langle|\psi_6|\rangle=1$ for a perfect system with hexagonal symmetry, 
$\langle|\psi_6|\rangle\simeq 0.3$ for stripes phases, and $\langle|\psi_6|\rangle=0$ for an isotropic disordered system in the thermodynamic limit.

We observe $\langle|\psi_6|\rangle$ values of ranging from $\simeq 0.3$ (stripes phase) to $1$ (triangular phase), with intermediate values $\langle|\psi_6|\rangle\simeq 0.5$ for coexisting phases (Fig.~\ref{fig:H_vs_rho_ALL}).

The $H$-index tends to decrease exponentially from anti-hyperuniform to nearly-hyperuniform values as density increases (dashed black line in Fig.~\ref{fig:H_vs_rho_ALL}).
This behavior is characteristic of particles that order upon an increase in density, thereby suppressing large-scale density fluctuations 
and reducing $H$.

The most compact layers, which form triangular lattices with minimal defects ($\langle|\psi_6|\rangle \simeq 1$), exhibit the lowest 
$H\simeq 9 \times 10^{-4}$, similar to nearly hyperuniform systems.
However, for all layers with $\rho \simeq 0.5$, regardless of the overall system $\rho$ and $T$, we observe $\langle|\psi_6|\rangle \simeq 0.3$. 
These layers exhibit stripe patterns, and their $H$ decreases from approximately $3 \times 10^{-1}$ to $2 \times 10^{-3}$, indicating the emergence of hyperuniformity.

To differentiate between fluid- and solid-like layers, we compute the mean square displacement (MSD) in 2D, defined as $\langle(\Delta r_{xy})^2(t)\rangle\equiv\langle(r_{xy}(t)-r_{xy}(0))^2\rangle$, averaged within a layer. Layers are considered diffusive if $\langle(\Delta r_{xy})^2(t_0)\rangle>R_0^2$, where $R_0^2$ the average area per particle.

\begin{figure}
\begin{center}
\includegraphics[clip=true,width=8.5cm]{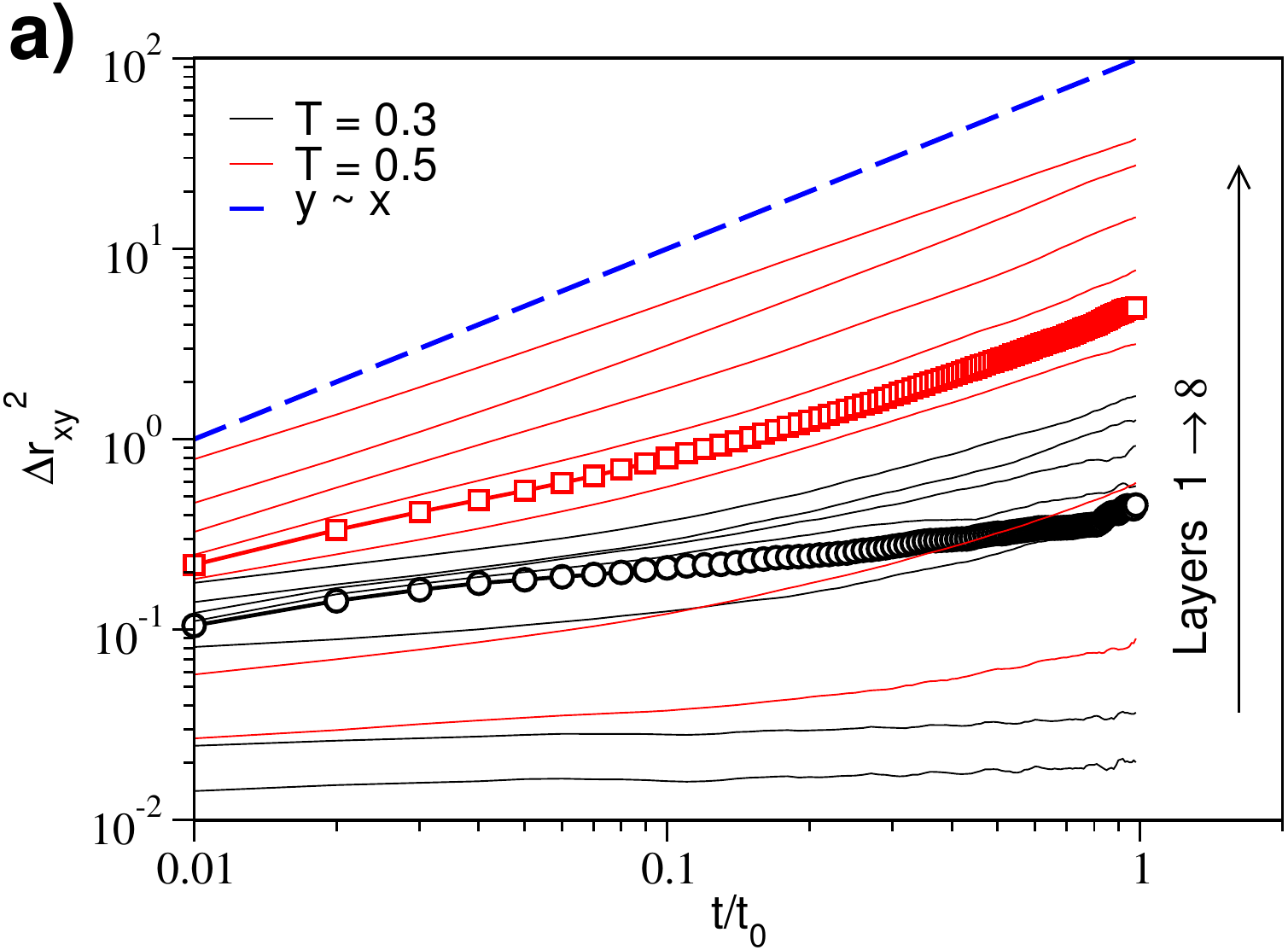} 
\includegraphics[clip=true,width=4.52cm]{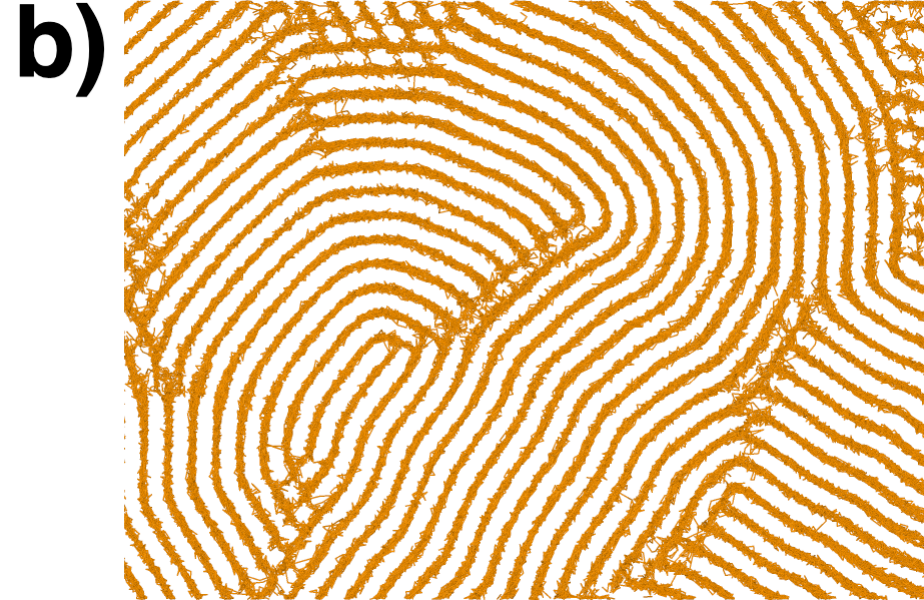}
\includegraphics[clip=true,width=3.95cm]{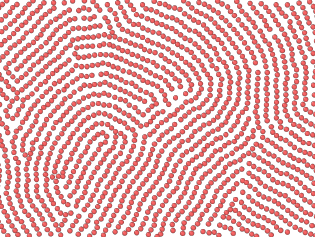}
\end{center}
\caption{{\bf Solid-like and fluid-like layers.}
(a) 2D MSD for $\tau=t_0$ at $\rho=0.3$ and $T=0.3$ (black) or $0T=.5$ (red) for $l_i$ with $i=1, \dots , 8$ from bottom to top.
At $T=0.3$ (black), all layers are solid-like, as indicated by $\langle(\Delta r_{xy})^2(t_0)\rangle<R_0^2\simeq 1.8$. 
At $T=0.5$ (red), layers $l_i$ with $i\geq 3$ exceed the $1.8$ threshold, consistent with fluid-like, diffusive behavior (blue dashed line). 
Lines with symbols represent $l_4$, having a stripe structure at both temperatures.
(b) At $T=0.5$, the stripe phase in $l_4$ (right panel, showing a small portion of the $xy$ snapshot at $t=0$) evolves along fluid-like trajectories 
(wiggly lines in the left panel) over a time window $t_0$, while preserving the stripe pattern. 
}
\label{fig:MSD}
\end{figure}

For fast annealing at $\rho=0.3$, we observe that all layers at $T=0.3$ are solid-like. However, at $T=0.5$, only $l_1$ and $l_2$,remain solid-like,  while the the others layers become fluid-like (Fig.~\ref{fig:MSD}).
Interestingly, fluid-like layers at $T=0.5$ have similar (or identical) values of $H$ values to those of corresponding solid-like layers 
at $T=0.3$ (Fig.~\ref{fig:H-index}, lower left panel), indicating that the fluidity of the structure does not correlate to its hyperuniform or anti-hyperuniform characteristics.

To investigate this behavior, we examine the particle trajectories in $l_4$ at $T=0.5$ (Fig.~\ref{fig:MSD}b). We observe that particles stochastically diffuse within stripe channels while maintaining their structure with a nearly hyperuniform $H \approx 3 \times 10^{-3}$.

%
%

\vspace{0.5cm}
\noindent{\bf Conclusion and outlook}

We numreically investigate the suppression of large-scale density fluctuations in self-assembled layers induced by slit-pore confinement of the CSW model, which applies to anomalous fluids such as globular proteins, soft colloids, liquid metals, and, to a certain extent, water. By varying thermodynamic conditions and slit-pore width, we facilitate the formation of layered structures exhibiting different types of long-range ordering, identifying nearly hyperuniform, non-hyperuniform, and anti-hyperuniform structures.
In particular, we observe 
1. nearly hyperuniform layers corresponding to 
a) crystals with point defects and grain boundaries, or 
b) disordered labyrinth-like stripe patterns; 
2. non-hyperuniform layers composed of coexisting crystalline patches with varying symmetries and defects, indicative of phase separation between different structures;
3. anti-hyperuniform crystalline layers containing extended, uncorrelated void defects (cavitation).

While anti-hyperuniform structures often arise, such as in phase separation processes~\cite{torquato2018}, we find that disordered, labyrinth-like stripe patterns—which frequently exhibit hyperuniform or nearly hyperuniform behavior—are prevalent across various systems. These patterns have been identified in a broad range of contexts, including magnetic fluids, amphiphilic monolayers, type I superconductors \cite{Akiva1993}, block copolymer films \cite{feng2022}, as well as in biological and physical processes like brain convolution development \cite{Richman1975}, fingerprint formation \cite{efimenko2005}, and skin wrinkling \cite{cerda2002}, among others.

We note that the CSW potential serves as a generic model for systems exhibiting thermodynamic, dynamic, and structural anomalies, similar to those seen in water. Consequently, our results may have broader implications for the category of anomalous fluids.
For instance, a recent study on a widely used atomistic water model—and to a lesser extent, a coarse-grained water model—demonstrated that the hyperuniformity of amorphous ices can exhibit non-monotonic behavior as a function of pressure at a constant temperature \cite{gartner2021}. Our observation of a pronounced minimum in  $H$  at the density corresponding to the stripe phase (Fig.~\ref{fig:H_vs_rho_ALL}) suggests that the isothermal  pressure-dependent, non-monotonic behavior of hyperuniformity noted in Ref.~\cite{gartner2021} arises from structural anomalies that are common to both the CSW model and other anomalous fluids.

Interestingly, studies on a model of epithelial tissue show that density fluctuations in the rigid phase are only suppressed up to a finite length scale. This scale increases rapidly as the system nears the fluid phase, ultimately reaching a state of effective hyperuniformity \cite{D0SM00776E}. Together with our findings, this suggests that the transition from rigid to fluid states in tissue monolayers may involve subtle changes in their confinement. For example, in our data at  $\rho = 0.3$  and  $T = 0.5$, hyperuniform layers can exhibit either solid-like behavior ($l_1$) or fluid-like characteristics 
($l_3$ and $l_4$) depending on their confining state (Fig.~\ref{fig:MSD}). This offers a fresh perspective for exploring rigidity transitions in biological tissues.

In a two-dimensional system with a hard core surrounded by a soft corona potential, the stripe phase that appears across a range of densities is associated with a peak in the specific heat \cite{malescio2003}. We anticipate a similar outcome in the CSW model, despite its inclusion of an attractive term in the potential. While this term may alter phase stability, it is likely only under conditions of extreme confinement \cite{leoni2017}.
Compared to the pure 2D system studied in Ref.~\cite{malescio2003}, demonstrating specific heat peaks across a range of densities in our case is more challenging because the system naturally selects the density for each layer, whereas we can control only the total system density. Moreover, we expect the heat capacity to exhibit multiple peaks at distinct densities, each corresponding to the formation of different phases in individual layers.
This finding suggests that by adjusting confinement size, it may be possible to induce changes in the fluid’s structure, phase behavior, and hyperuniformity. We aim to explore these confinement-induced effects on fluids in future studies.

\begin{acknowledgments}
We thank S. Torquato and A. Gabrielli for their valuable discussions, and we extend our sincere gratitude specifically to Fausto Martelli for the insightful discussions that greatly contributed to this study. 
E. C. O acknowledges support from RFIS-II Grant by the National Natural Science Foundation of China (Grant No. 12350610238). 
G.F. is grateful for support by MCIN/AEI/ 10.13039/ 501100011033 and “ERDF A way of making Europe” grant number PID2021-124297NB-C31, by the Ministry of Universities 2023-2024 Mobility Subprogram within the Talent and its Employability Promotion State Program (PEICTI 2021-2023), and by the Visitor Program of the Max Planck Institute for The Physics of Complex Systems for supporting a visit started in November 2022. 
\end{acknowledgments}

\bibliography{biblio}

\setcounter{equation}{0}
\setcounter{figure}{0}
\setcounter{table}{0}
\setcounter{section}{0}
\makeatletter
\renewcommand{\theequation}{S\arabic{equation}}
\renewcommand{\thefigure}{S\arabic{figure}}
\renewcommand{\thetable}{S\arabic{table}}
\renewcommand{\thesection}{S\arabic{section}}

\clearpage
\newpage
\onecolumngrid

\begin{center}
{\Large\bf{Supplementary Material for ``Emergence of Disordered Hyperuniformity in Confined Fluids and Soft Matter''}}
\end{center}
\vspace{0.3cm}

\section{Density profiles}

\begin{figure}[h!]
\includegraphics[clip=true,width=8cm]{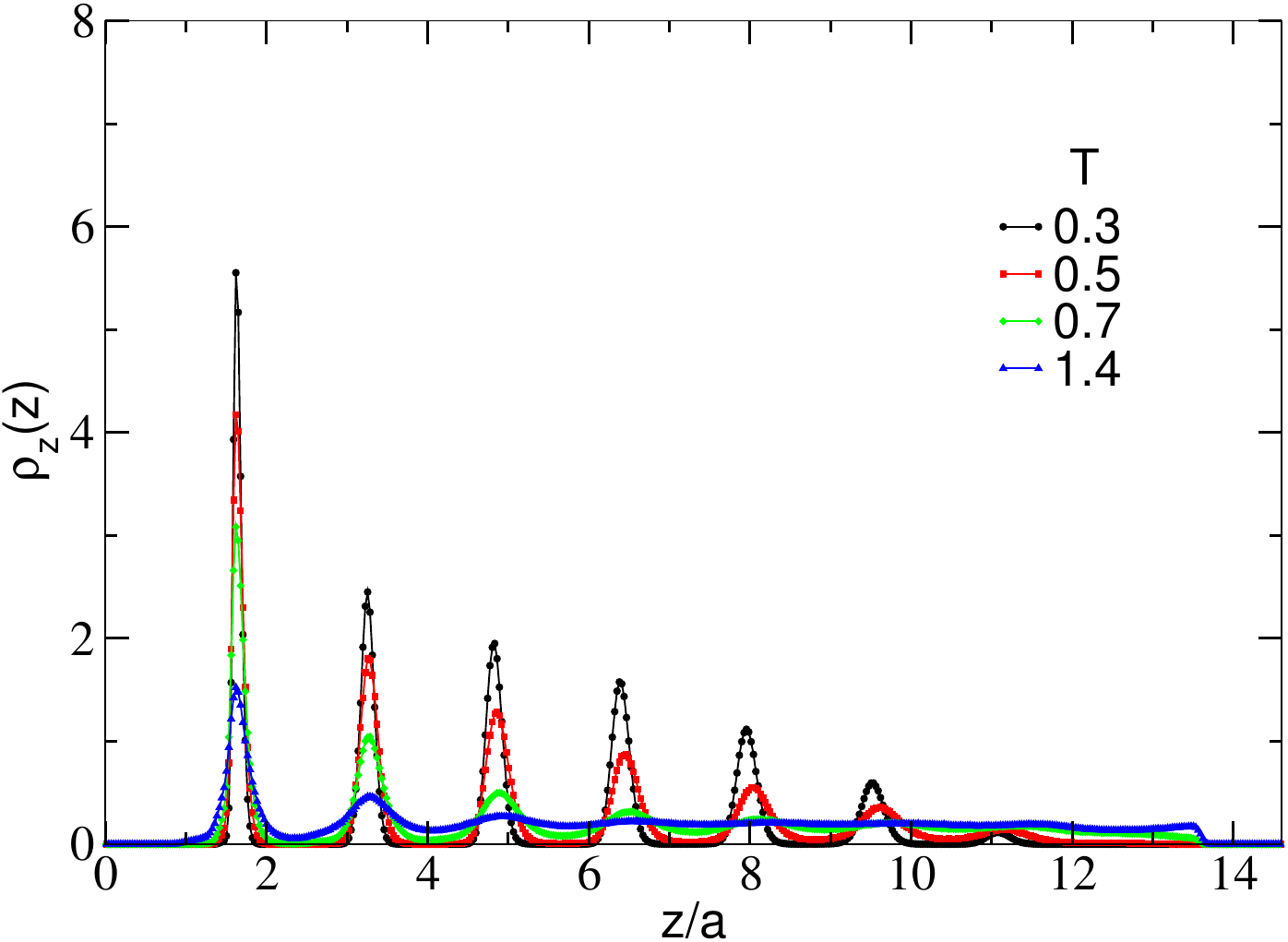}
\hspace{0.25cm}
\includegraphics[clip=true,width=8cm]{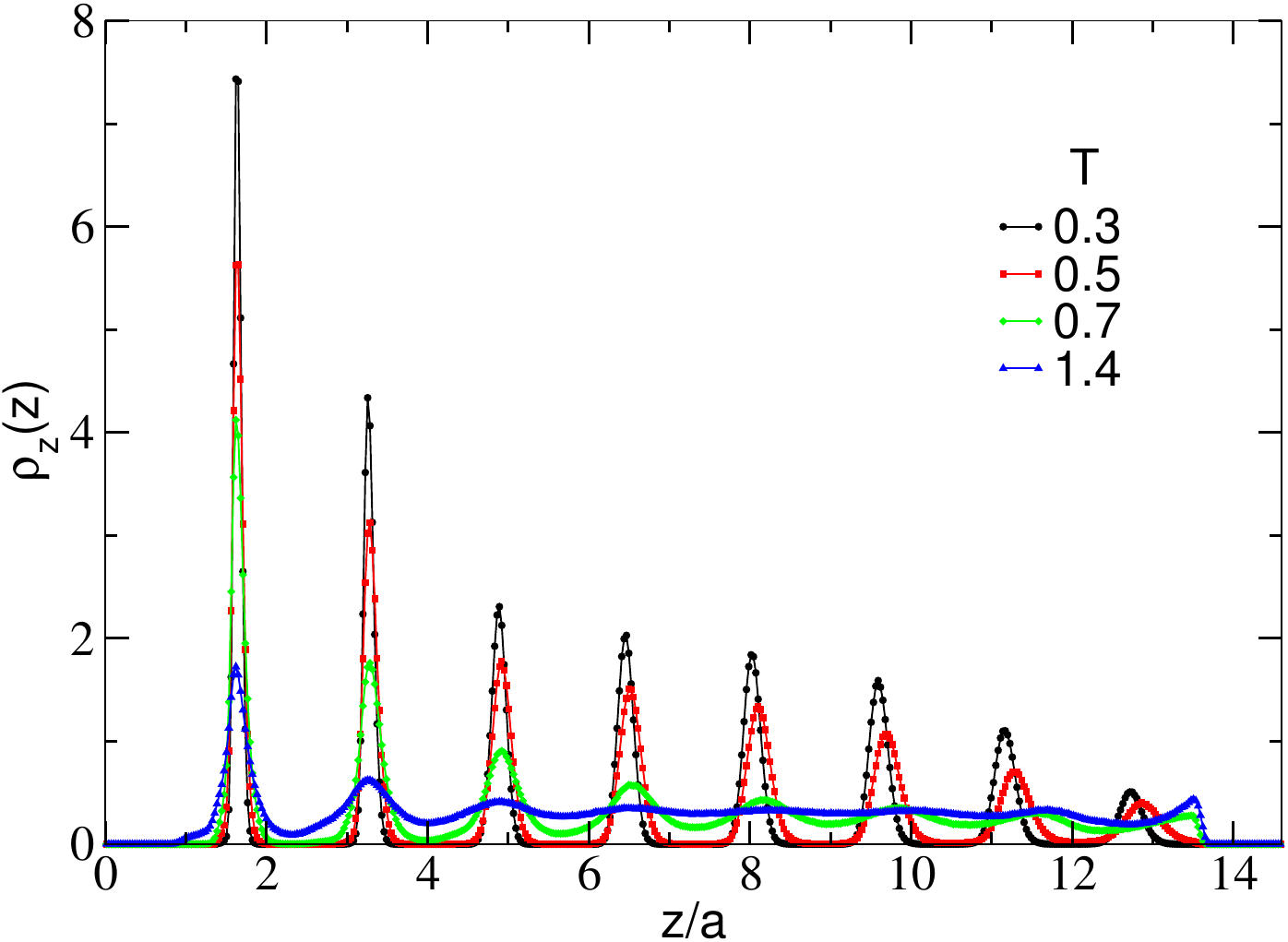}
\caption{Density profiles along the $z$-axis for the 8-layer system at fluid densities $\rho=0.2$ (left) and $\rho=0.3$ (right) at various temperatures under fast annealing conditions ($\tau=t_0$).}  
\label{figRHOz_rho0.2_T0.3_tau_t0}
\end{figure}

\begin{figure}[h!]
\includegraphics[clip=true,width=8cm]{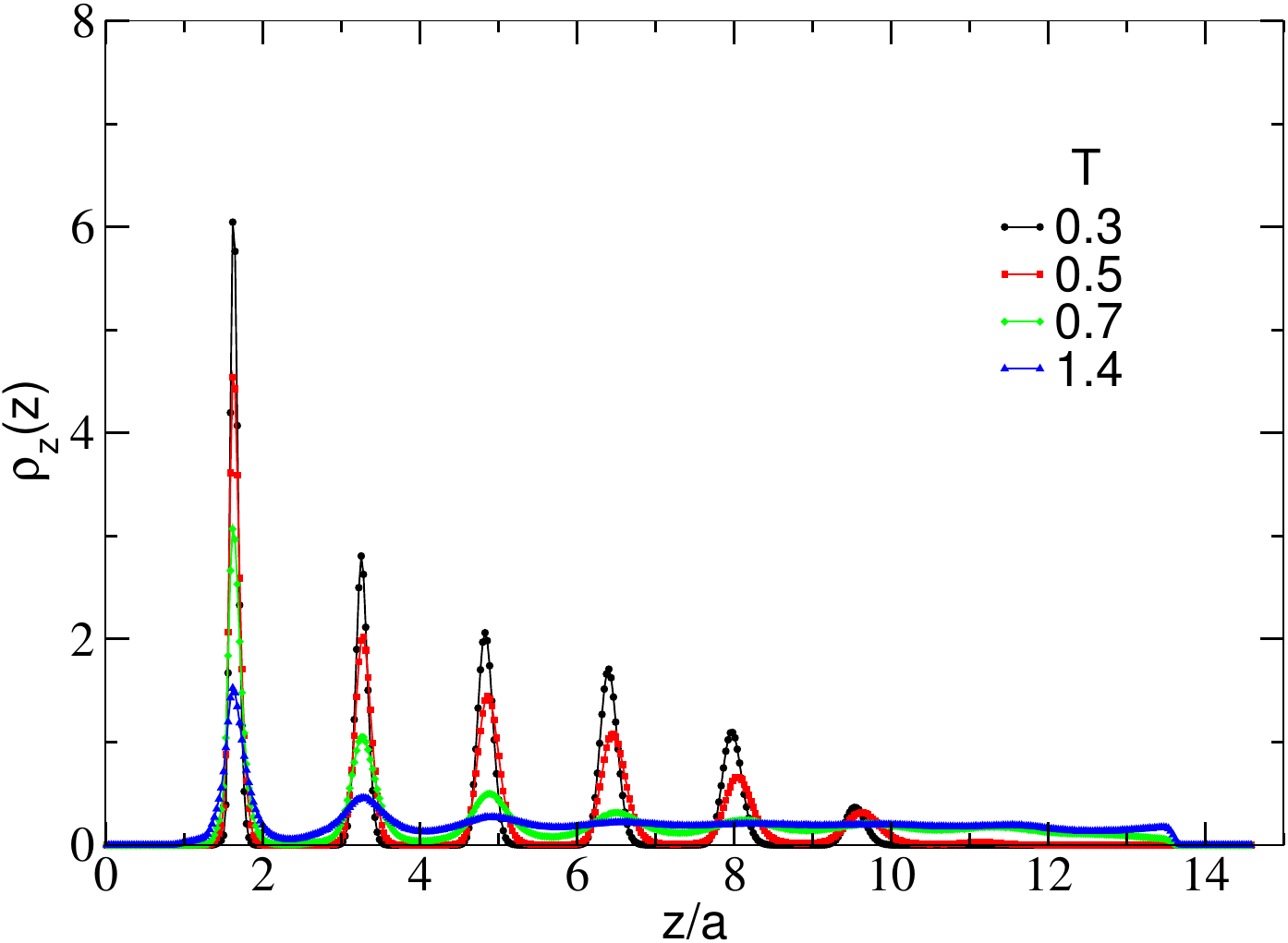}
\hspace{0.25cm}
\includegraphics[clip=true,width=8cm]{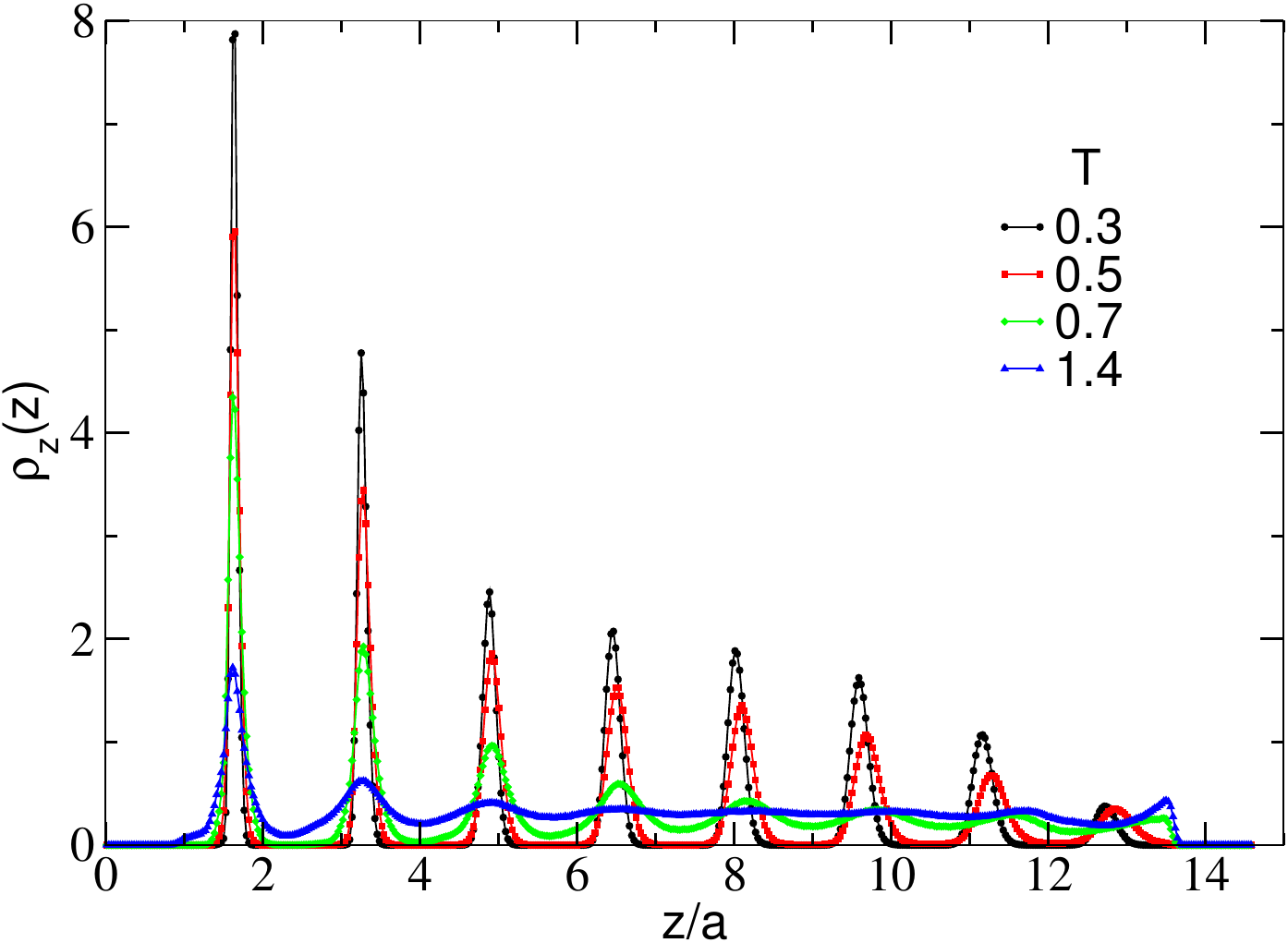}
\caption{
As in Fig. \ref{figRHOz_rho0.2_T0.3_tau_t0}, but for slow annealing ($\tau=10t_0$).}  
\label{figRHOz_rho0.2_T0.3_tau_10t0}
\end{figure}

\newpage
\section{Configurations}
\begin{figure}[h!]
\includegraphics[clip=true,width=3.1cm]{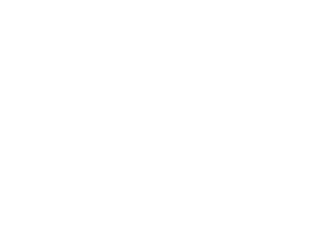}
\includegraphics[clip=true,width=3.1cm]{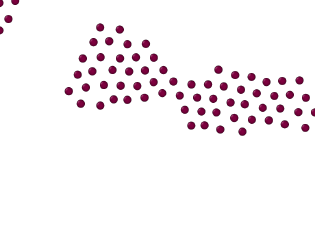}
\hspace{1cm}\includegraphics[clip=true,width=3.1cm]{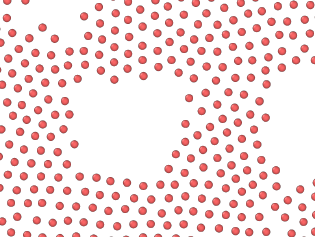}
\includegraphics[clip=true,width=3.1cm]{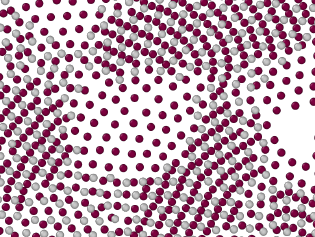}

\includegraphics[clip=true,width=3.1cm]{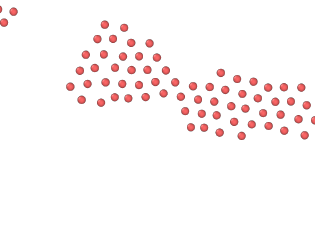}
\includegraphics[clip=true,width=3.1cm]{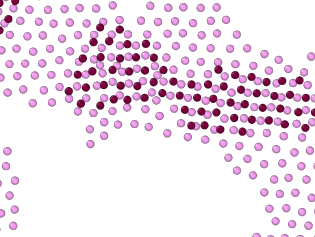}
\hspace{1cm}\includegraphics[clip=true,width=3.1cm]{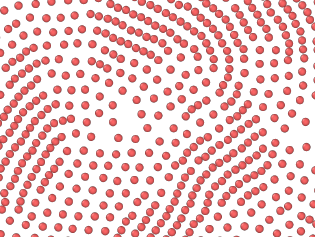}
\includegraphics[clip=true,width=3.1cm]{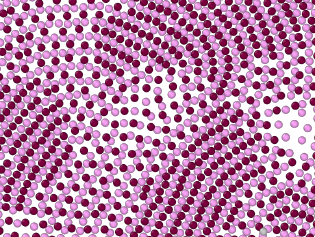}

\includegraphics[clip=true,width=3.1cm]{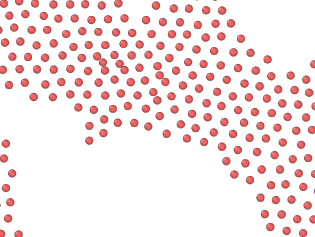}
\includegraphics[clip=true,width=3.1cm]{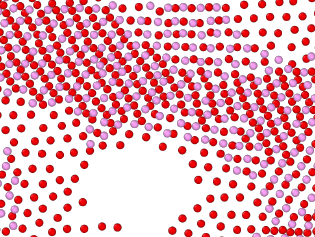}
\hspace{1cm}\includegraphics[clip=true,width=3.1cm]{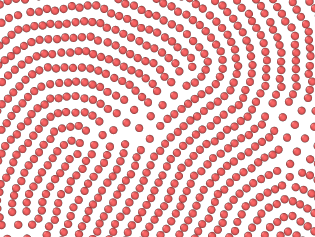}
\includegraphics[clip=true,width=3.1cm]{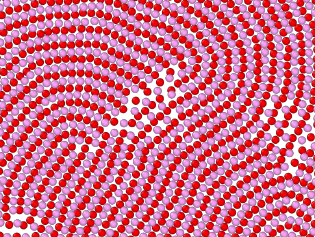}

\includegraphics[clip=true,width=3.1cm]{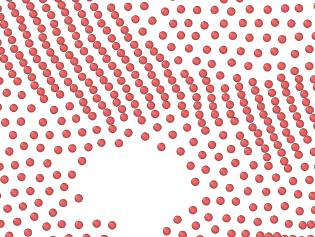}
\includegraphics[clip=true,width=3.1cm]{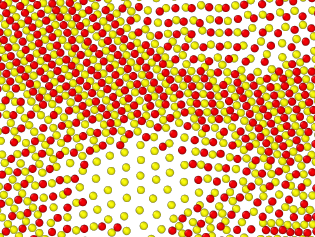}
\hspace{1cm}\includegraphics[clip=true,width=3.1cm]{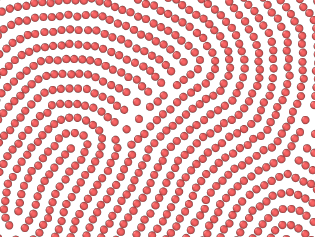}
\includegraphics[clip=true,width=3.1cm]{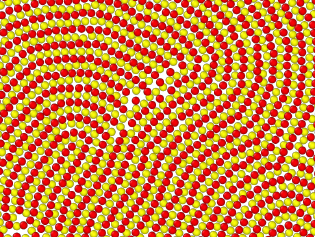}

\includegraphics[clip=true,width=3.1cm]{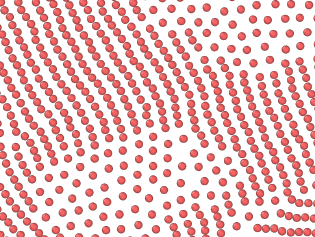}
\includegraphics[clip=true,width=3.1cm]{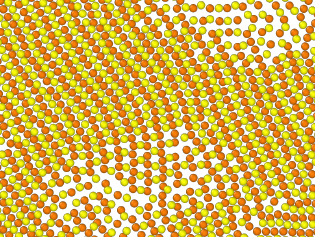}
\hspace{1cm}\includegraphics[clip=true,width=3.1cm]{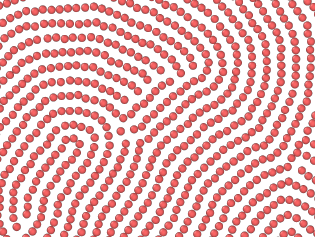}
\includegraphics[clip=true,width=3.1cm]{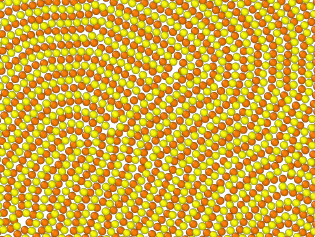}

\includegraphics[clip=true,width=3.1cm]{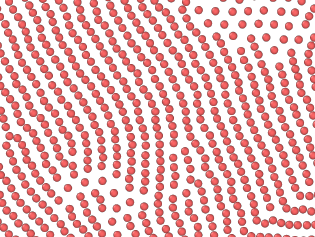}
\includegraphics[clip=true,width=3.1cm]{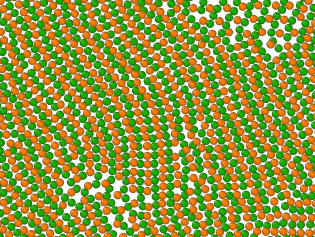}
\hspace{1cm}\includegraphics[clip=true,width=3.1cm]{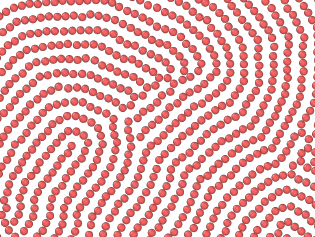}
\includegraphics[clip=true,width=3.1cm]{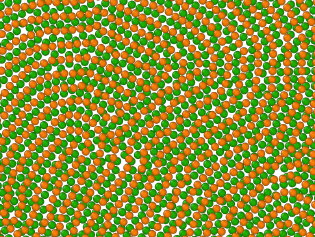}

\includegraphics[clip=true,width=3.1cm]{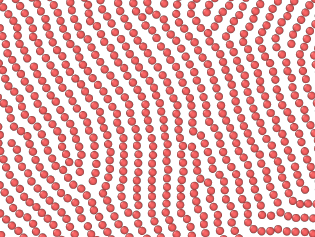}
\includegraphics[clip=true,width=3.1cm]{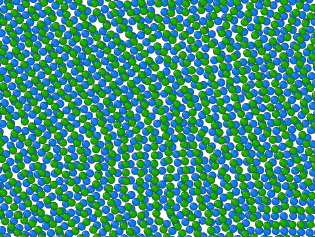}
\hspace{1cm}\includegraphics[clip=true,width=3.1cm]{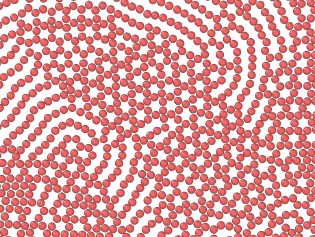}
\includegraphics[clip=true,width=3.1cm]{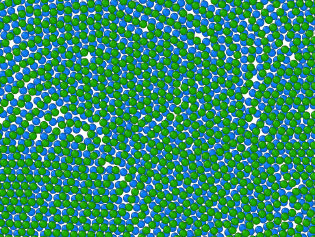}

\includegraphics[clip=true,width=3.1cm]{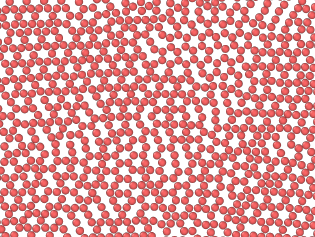}
\includegraphics[clip=true,width=3.1cm]{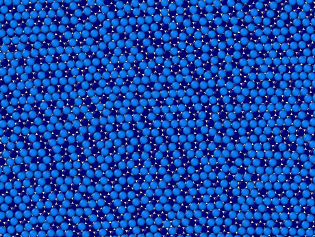}
\hspace{1cm}\includegraphics[clip=true,width=3.1cm]{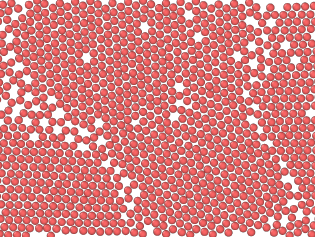}
\includegraphics[clip=true,width=3.1cm]{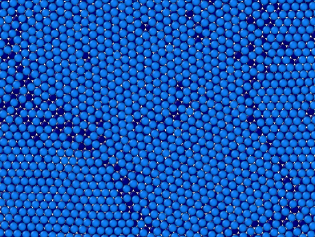}
\caption{As in Fig.~1 of the main text, showing layers $l_1$  to $l_8$ (from bottom to top panels) at $T=0.3$ with densities $\rho=0.2$ (first column from the left) and 
$\rho=0.3$ (third column from the left). The second and fourth columns from the left display the superposition of two adjacent layers for $\rho=0.2$ and $0.3$, respectively.
}  
\label{figsnap_molding}
\end{figure}

\begin{figure*}[t!]
\begin{center}
\includegraphics[clip=true,width=8cm]{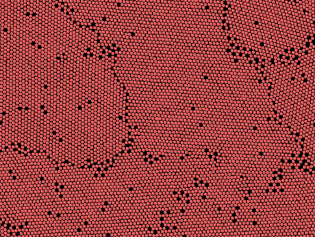}
\includegraphics[clip=true,width=8cm]{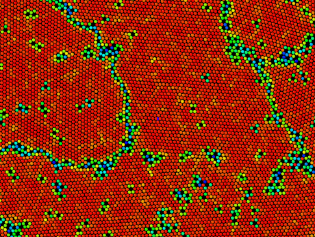}
\includegraphics[clip=true,width=8cm]{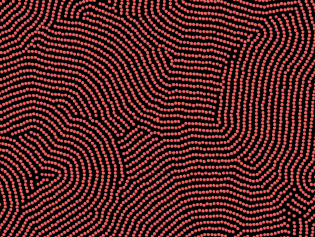}
\includegraphics[clip=true,width=8cm]{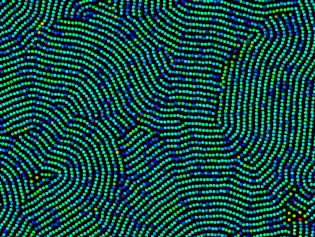}
\includegraphics[clip=true,width=8cm]{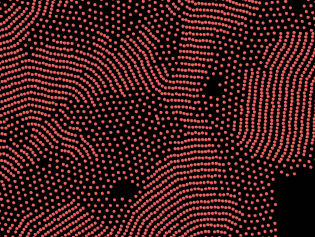}
\includegraphics[clip=true,width=8cm]{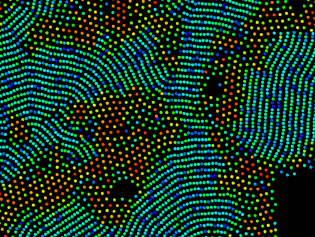}
\end{center}
\caption{Snapshots (left column) show a subregion of size 120x70 for layers $l_1$ (upper panels), $l_4$ (middle panels), and $l_7$ (bottom panels) at $\rho=0.3$, $T=0.3$, and $\tau=t_0$. The corresponding color maps (right column) indicate $|\psi_6|$ values, following the scale in Fig.~4
(from 0, blue, to 1, red).
Here, $|\psi_6|$ is calculated locally, with values ranging from approximately $0.3$ (stripe structure, turquoise) to $1$ (triangular structure, red), and intermediate values around $0.5$ (green and yellow) for defects.}
\label{figpsi6_rho_03_T03_l1}
\end{figure*}

\newpage\newpage\clearpage

\section{Quantifying hyperuniformity}

The variance $\sigma^{2}(R)$, associated with the number of particles within a circular observation window of radius $R$, can be written as 
\begin{equation}
 \sigma^{2}(R)\equiv \langle N(R)\rangle\left [\frac{1}{(2\pi)^d} \int_{\mathbb{R}^d}S(\bf{k})\tilde{\alpha}(\bf{k}; R)d\bf{k}\right]
 \label{eq:eq1}
\end{equation}
where $\langle N(R)\rangle$ represents the average number of particles in the observation window, $d$ the dimension of the Euclidean space, and $\tilde{\alpha}(\bf{k}; R)$ is the Fourier transform of the overlapping volume between two observation windows with centers separated by a vector $\bf{R}$, normalized by the volume of the windows. For many  ordered and disordered systems, the large-$R$ asymptotic behavior of Eq.~(\ref{eq:eq1}) can be expressed as~\cite{torquato2003local,martelli2017}
\begin{equation}
\sigma^2(R)=2^d\phi\left[A\left(\dfrac{R}{D}\right)^d+B\left(\dfrac{R}{D}\right)^{d-1}+O\left(\dfrac{R}{D}\right)^{d-1}\right],    
\label{AB}
\end{equation}
where $\phi$ is a dimensionless density~\cite{torquato2018,torquato2021b}, 
$D$ a characteristic microscopic length, 
$O(R/D)^{d-1}$ represents terms of order higher than $(R/D)^{d-1}$,
$A$ and $B$ are {\it volume} and {\it surface} coefficients, respectively, and are computed as \cite{torquato2021b}
\begin{equation}
A\equiv \lim_{|{\bm k}|\rightarrow 0}S({\bm k}),
\label{A}
\end{equation} 
and 
\begin{equation}
B\equiv d/\pi\int_{0}^{\infty}(S(k)-S(0))k^{-2} dk.
\label{B}
\end{equation}
%

\begin{figure}[t]
\includegraphics[clip=true,width=8cm]{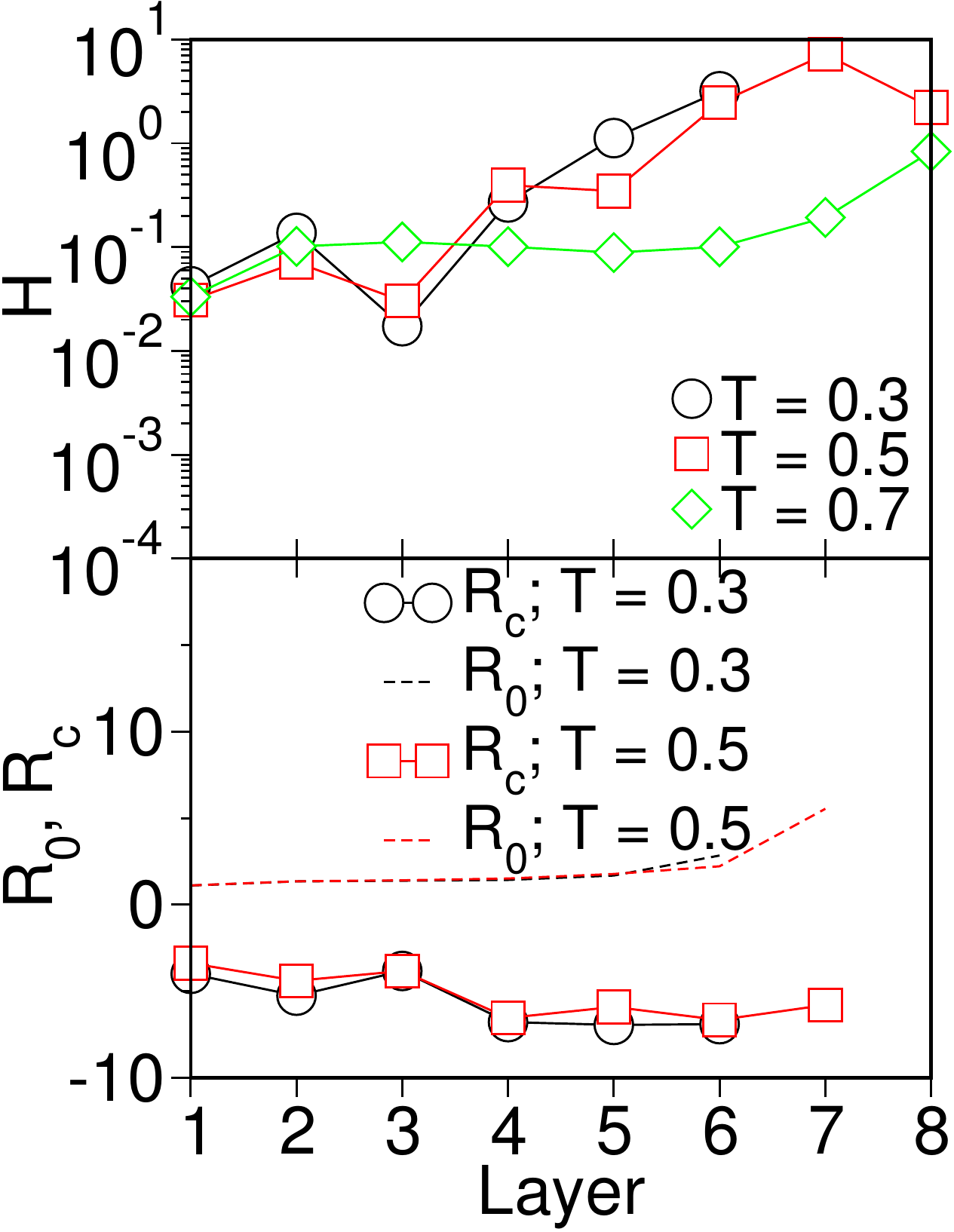} 
\includegraphics[clip=true,width=8cm]{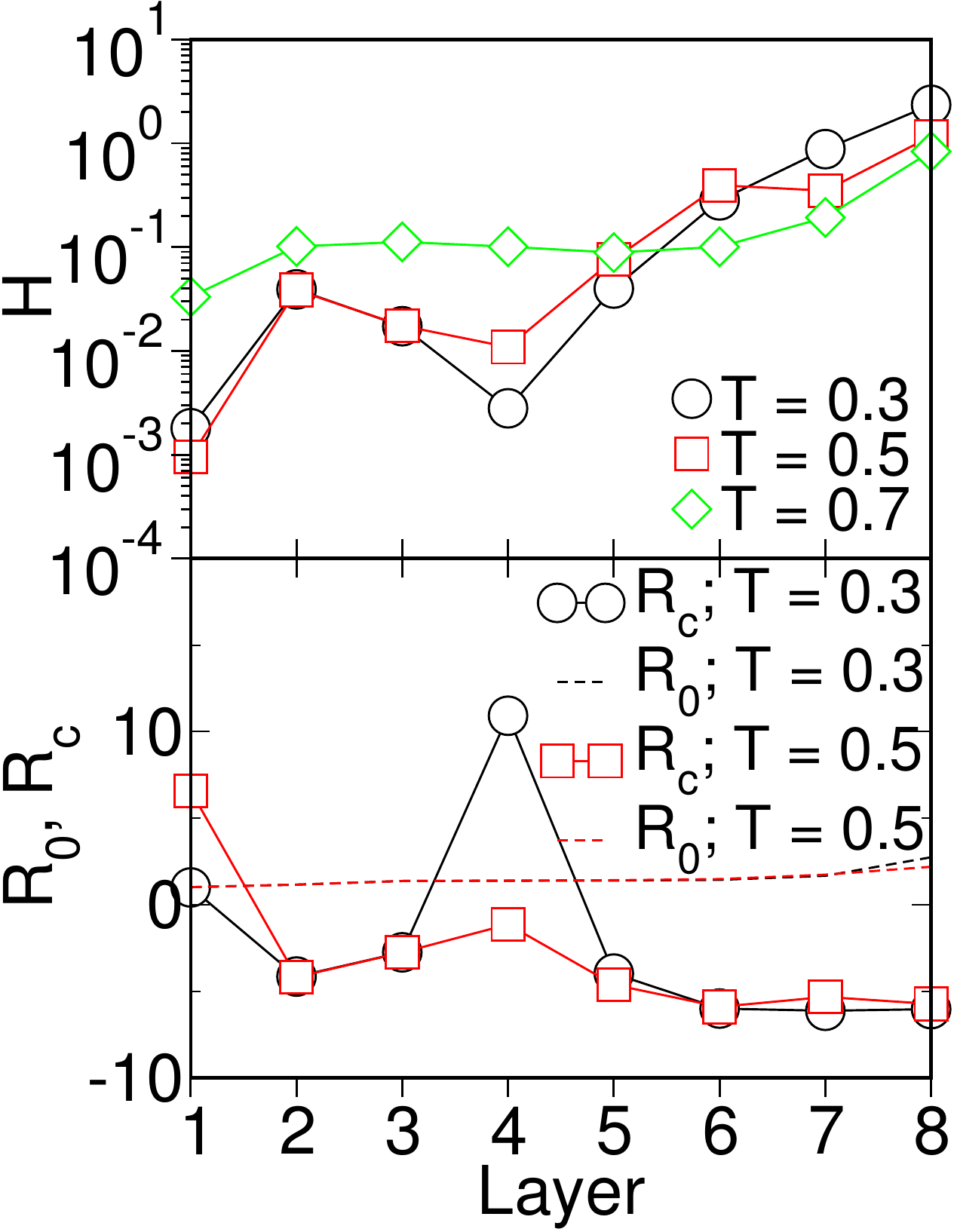}
\caption{As in Fig.3
of the main text, but for fast annealing ($\tau=t_0$). 
The upper panels the $H$ index, while the lower panels display the average nearest-neighbor distance $R_0\equiv \rho^{-1/d}$ (dashed lines) and $R_c\approx B/A$ (symbols), where $A$ and $B$ are defined by Eqs.(\ref{A}, \ref{B}), plotted against the layer number. 
The left panels represent $\rho=0.2$ and right panels  $\rho=0.3$ at $T=0.3$ (black), 0.5 (red).}
\label{figH-index_SM}
\end{figure}

\begin{figure*}[t!]
\begin{center}
\includegraphics[clip=true,width=8cm]{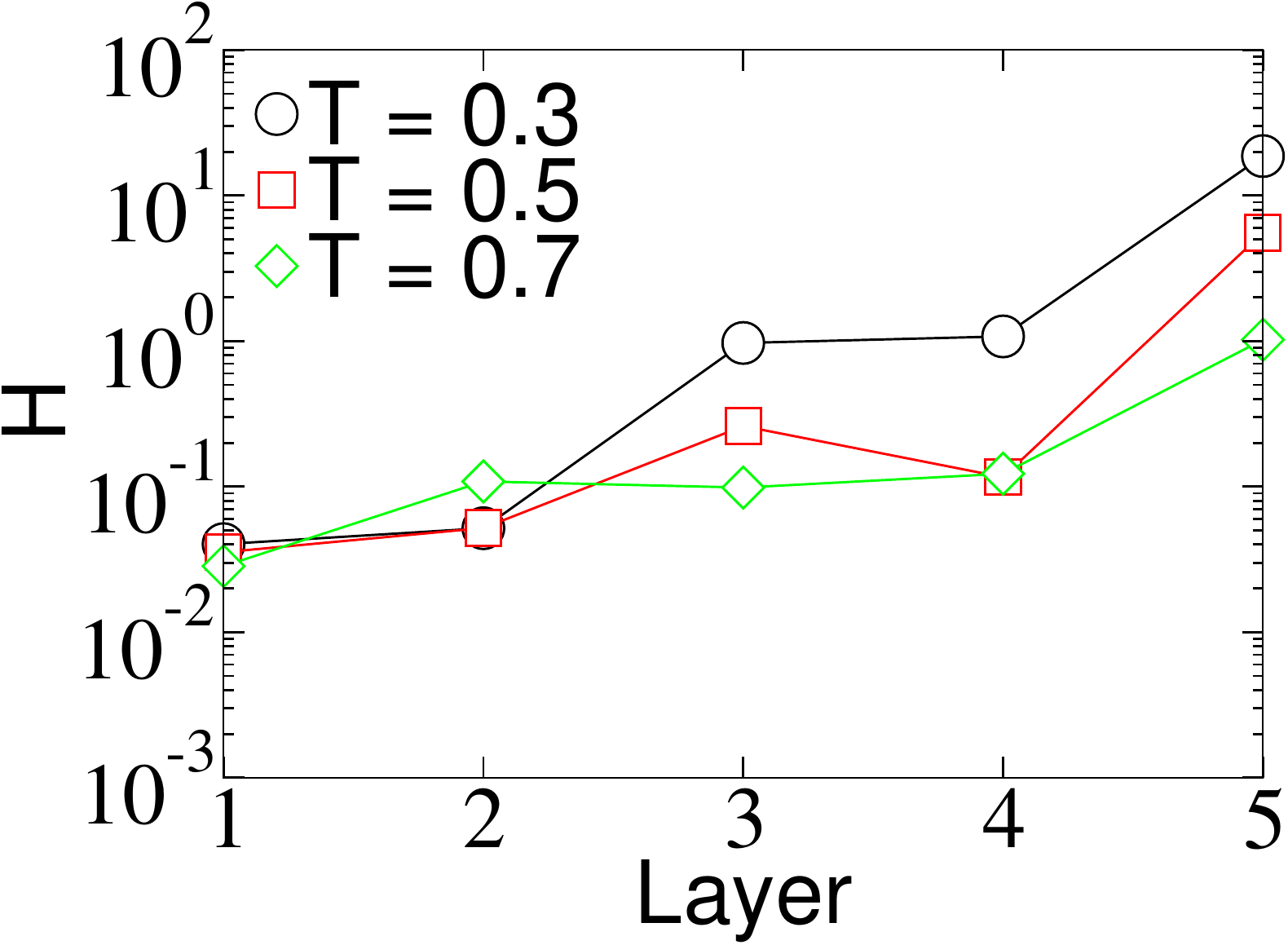} 
\includegraphics[clip=true,width=8cm]{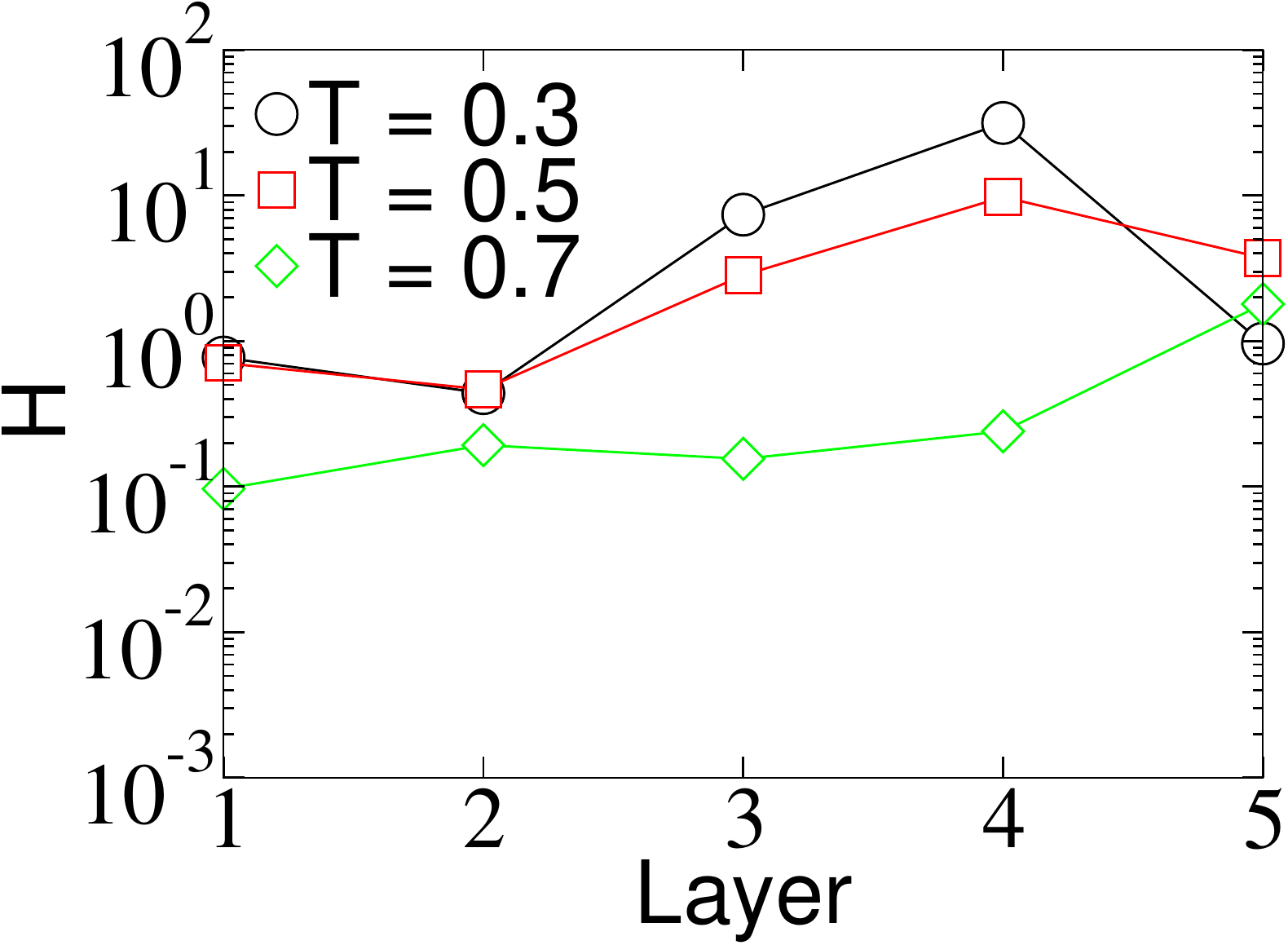} 

\includegraphics[clip=true,width=8cm]{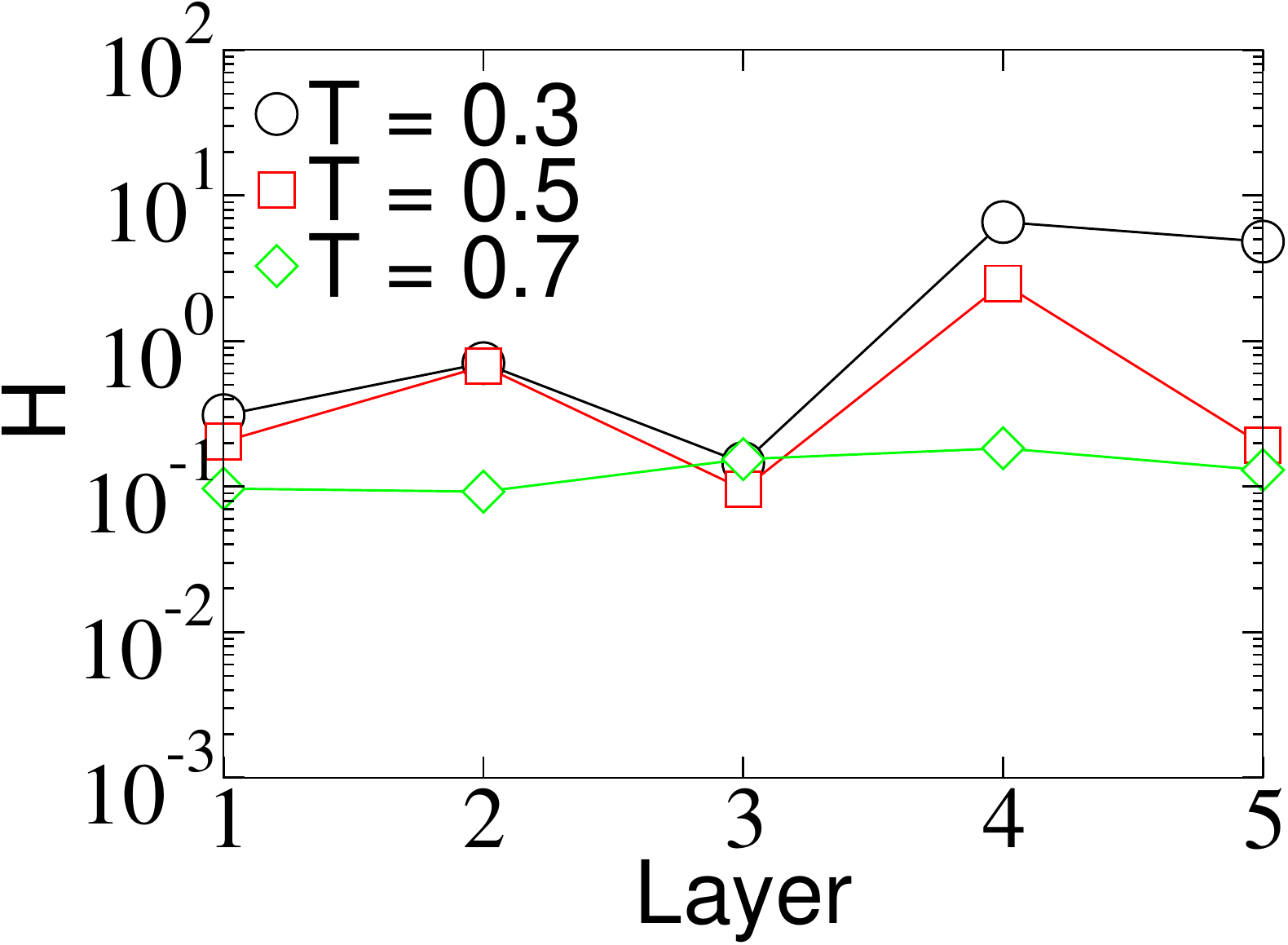} 
\includegraphics[clip=true,width=8cm]{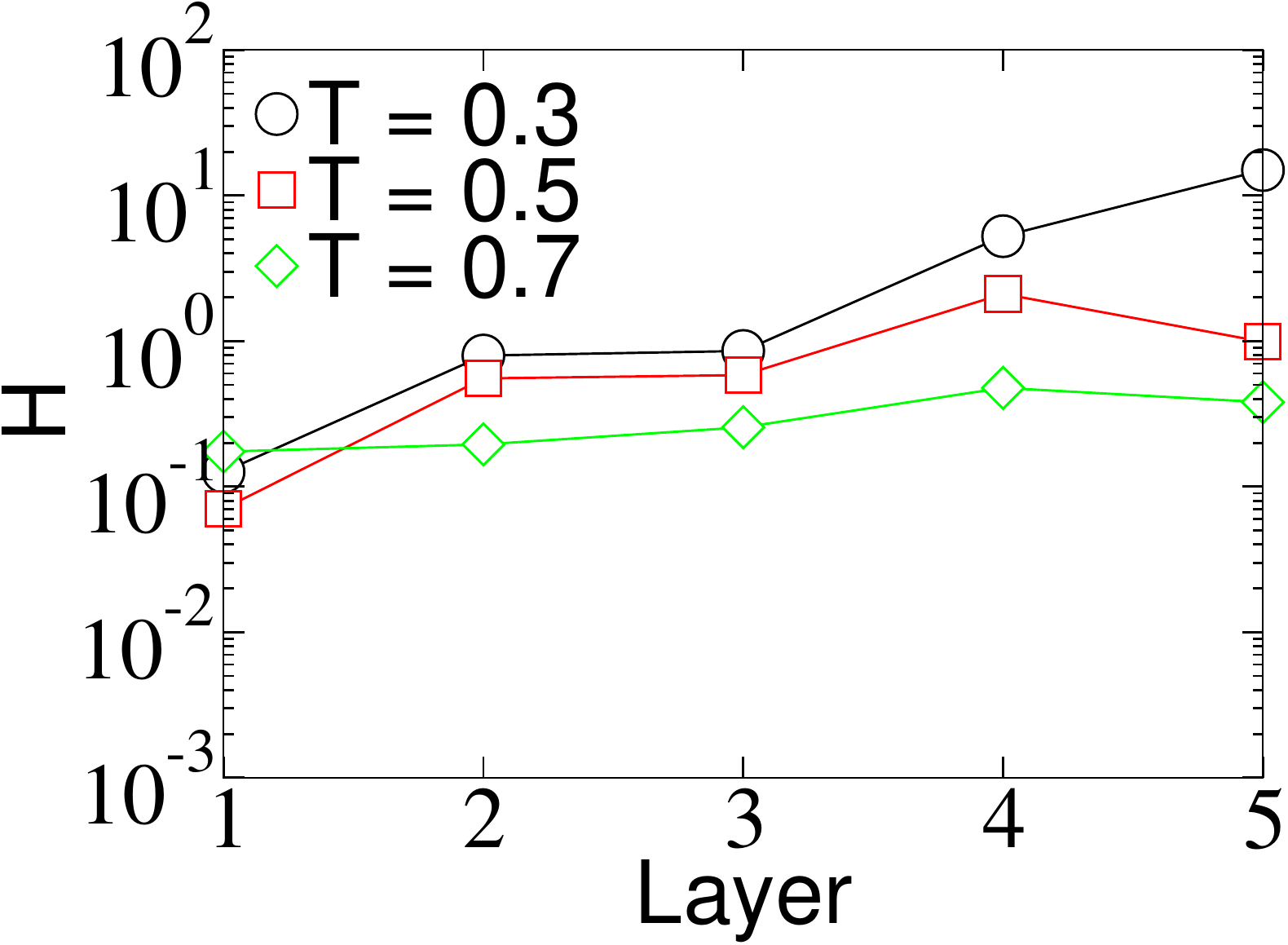} 
\end{center}
\caption{H-index for the 5-layer system. Upper panels correspond to $\rho=0.2$, while the lower panels correspond to $\rho=0.3$. Left panels represent $\tau=t_0$, right panels represent $\tau=10 t_0$.}   
\label{figH-index5}
\end{figure*}
\begin{figure*}[h!]
\begin{center}
\includegraphics[clip=true,width=8cm]{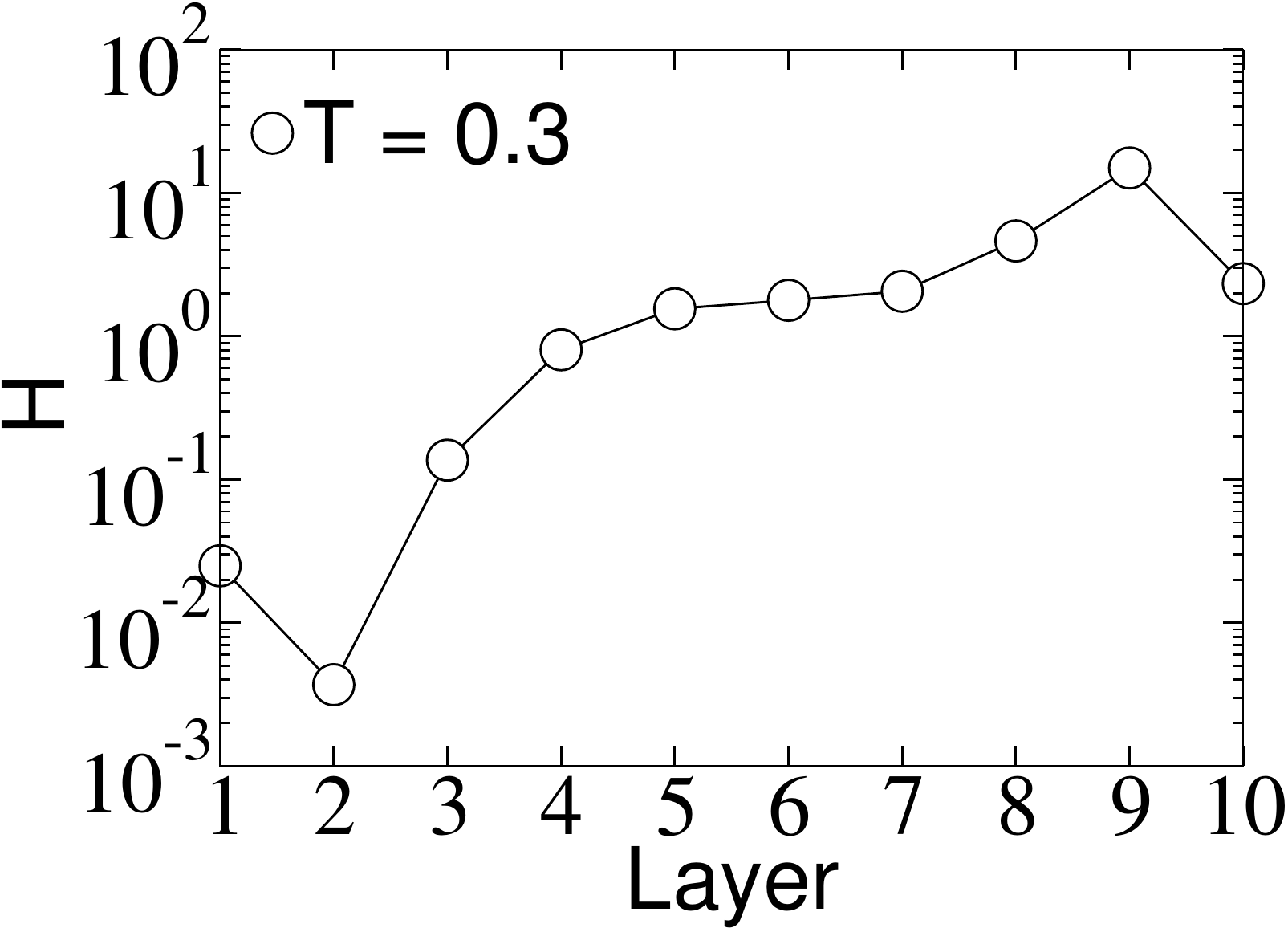} 
\includegraphics[clip=true,width=8cm]{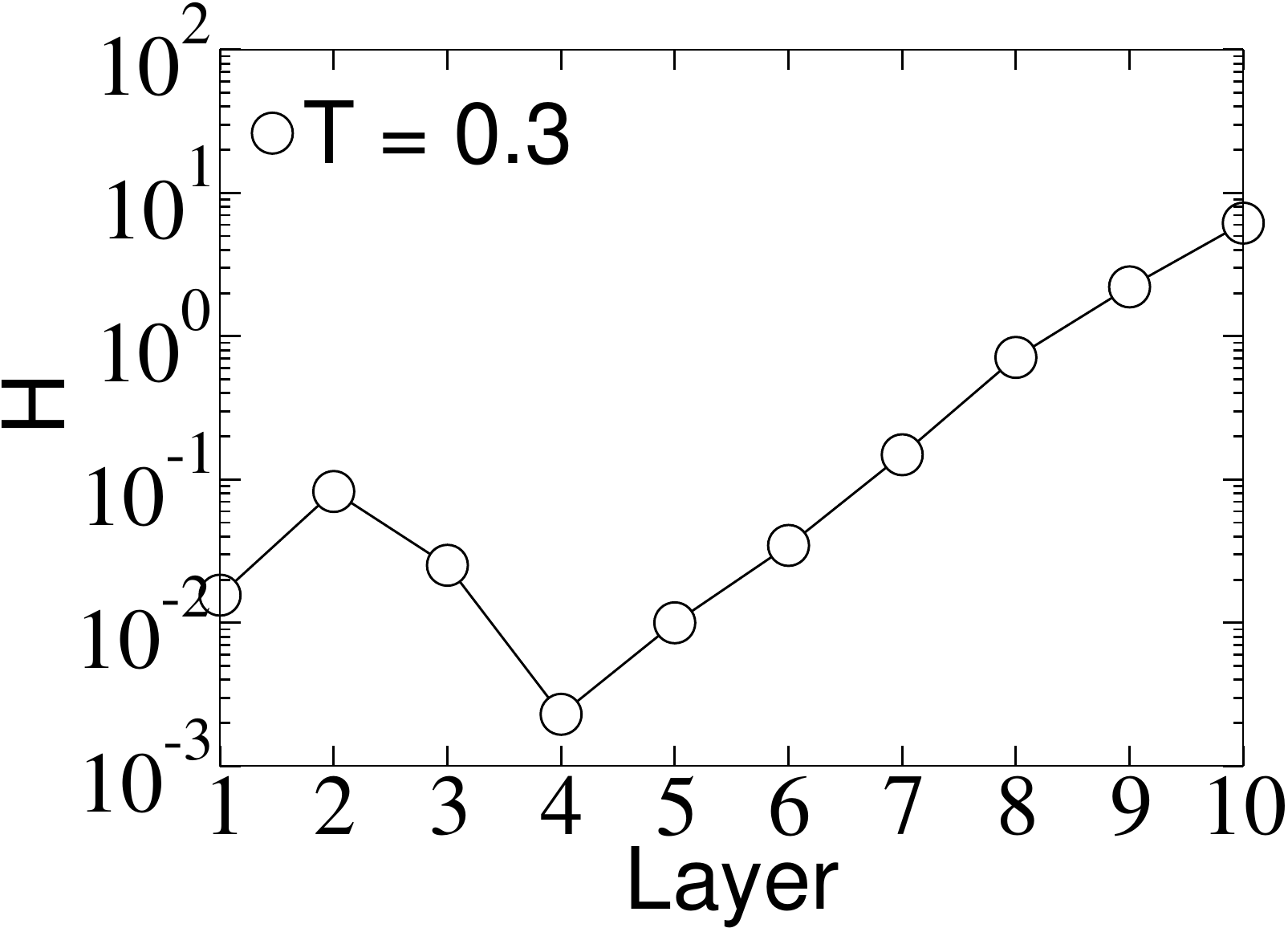} 
\end{center}
\caption{H-index for the 10-layer system at $T=0.3$ and $\tau=t_0$. Left panel shows results for $\rho=0.2$, and the right panel for $\rho=0.3$.} 
\label{figH-index10}
\end{figure*}
%

\begin{figure*}[h!]
\begin{center}
\includegraphics[clip=true,width=8cm]{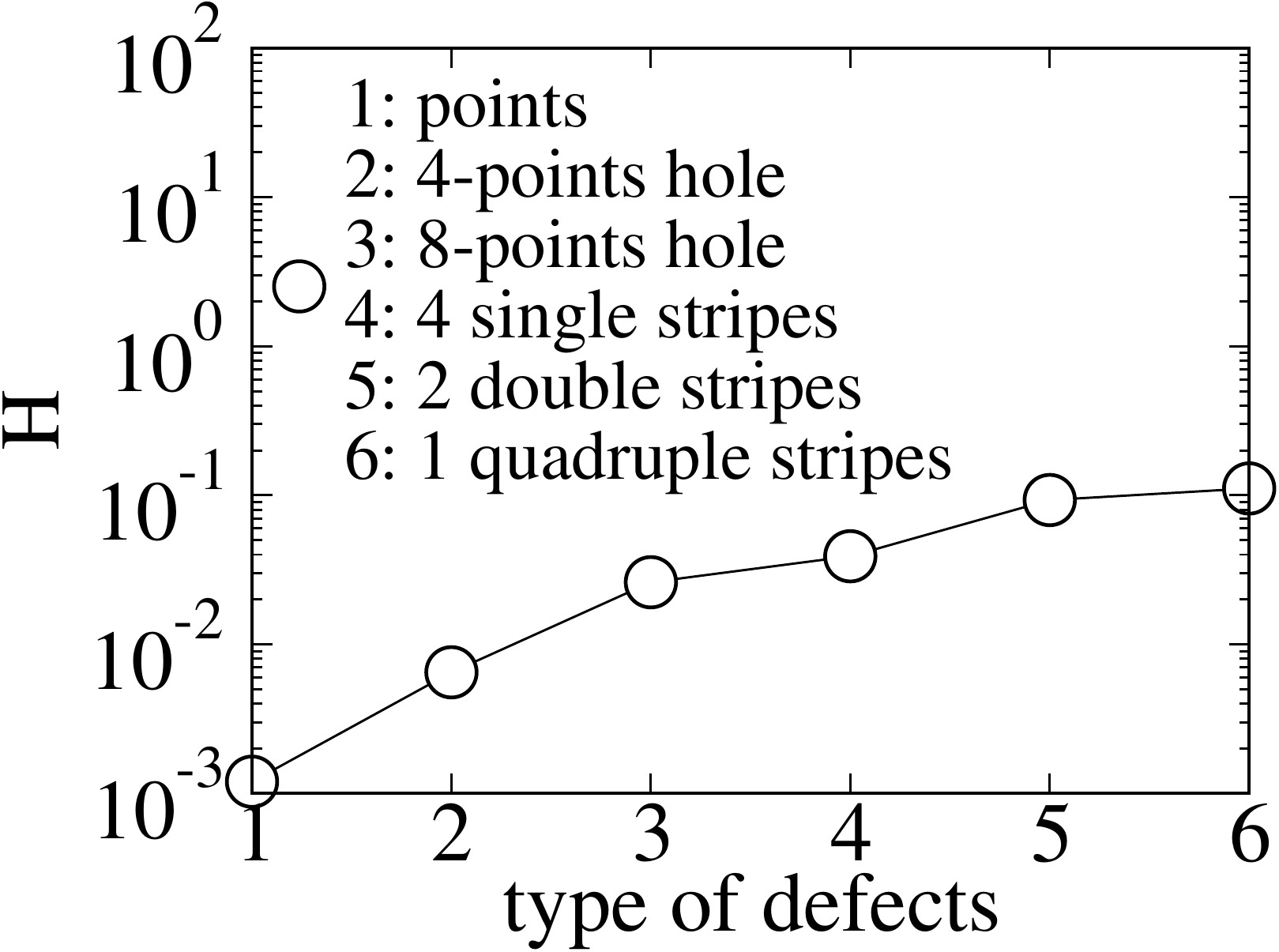}
\end{center}
\caption{H-index for the 2D perfect triangular lattice with different types of defects, at a fixed particle defect  concentration of $2\%$ across the entire system. Defect types are defined in Fig.~\ref{figsnapshots_defects}.}   
\label{figH_defects}
\end{figure*}
%

\begin{figure*}[h!]
\begin{center}
(1)
\includegraphics[clip=true,width=4.5cm]{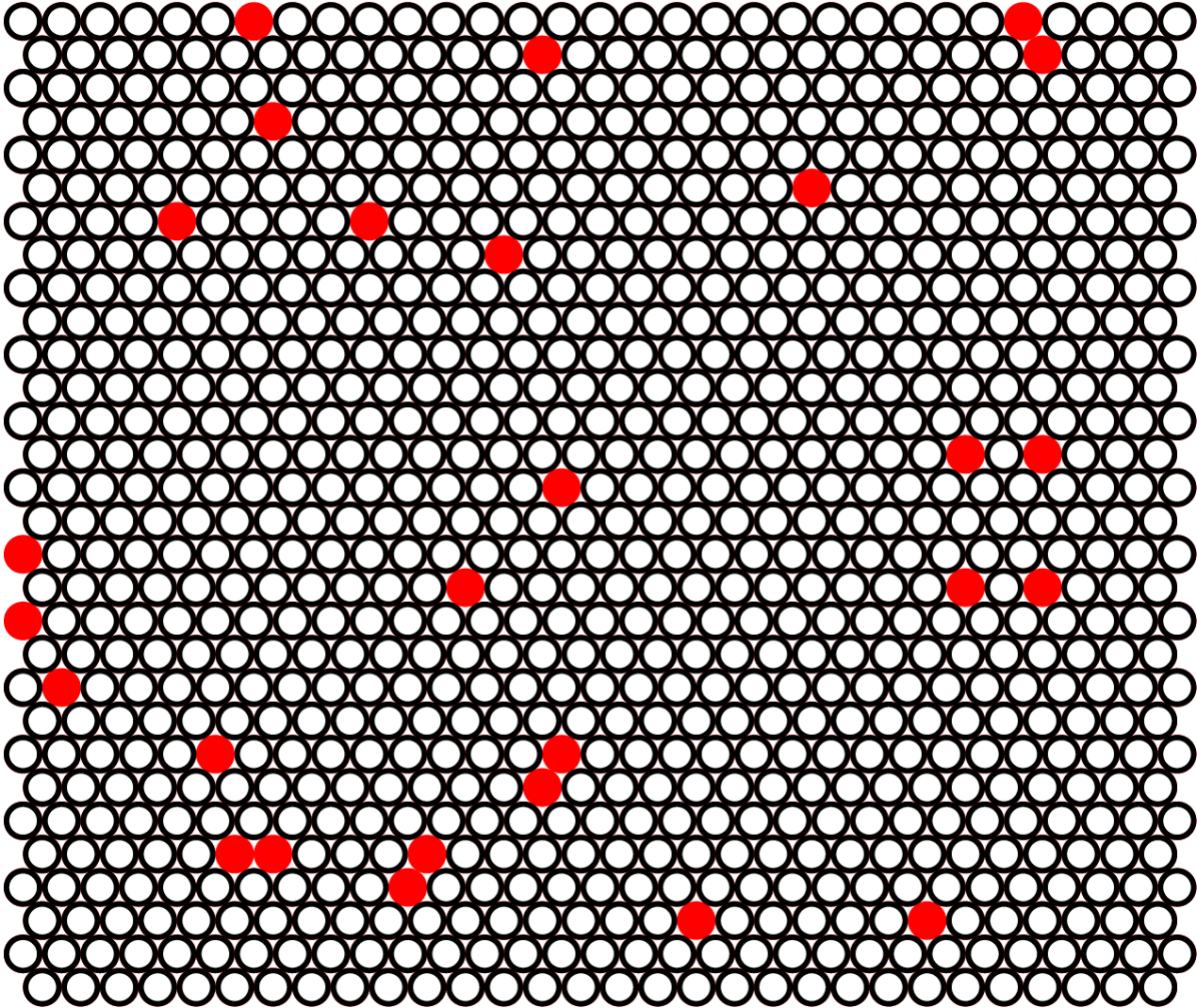}
\hspace{0.25cm}
(2)
\includegraphics[clip=true,width=4.5cm]{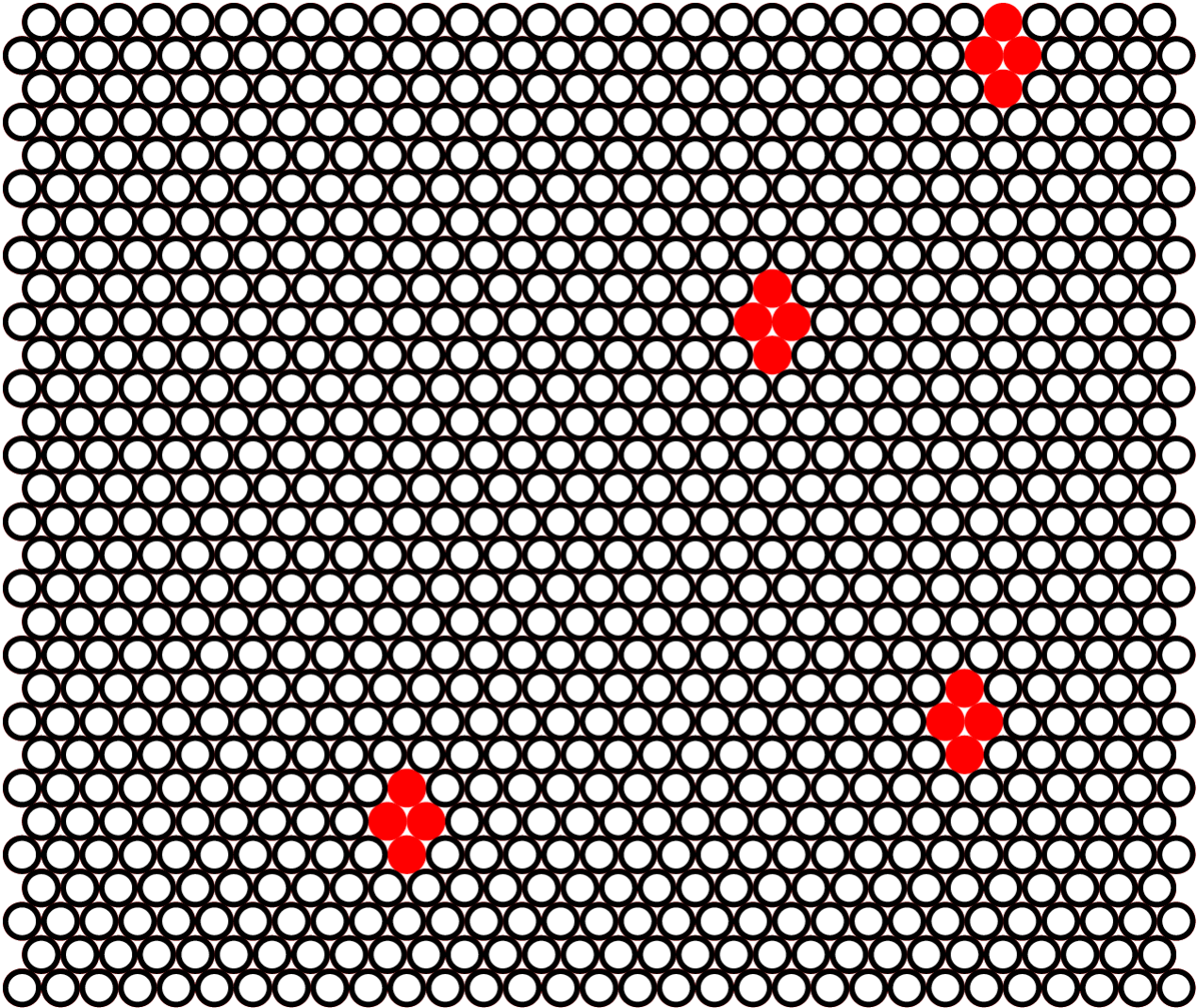}
\hspace{0.25cm}
(3)
\includegraphics[clip=true,width=4.5cm]{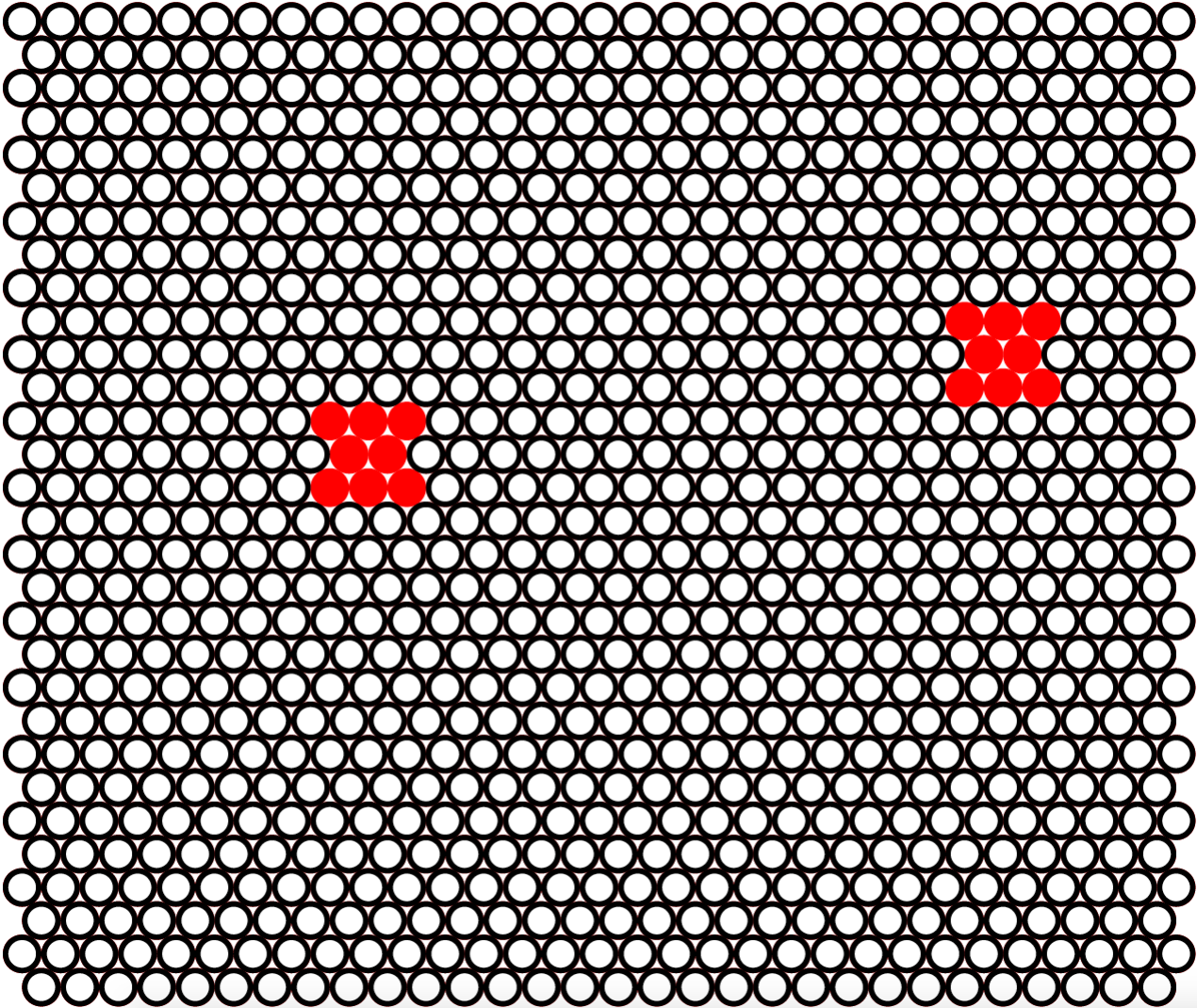}

\vspace{0.25cm}

(4)
\includegraphics[clip=true,width=4.5cm]{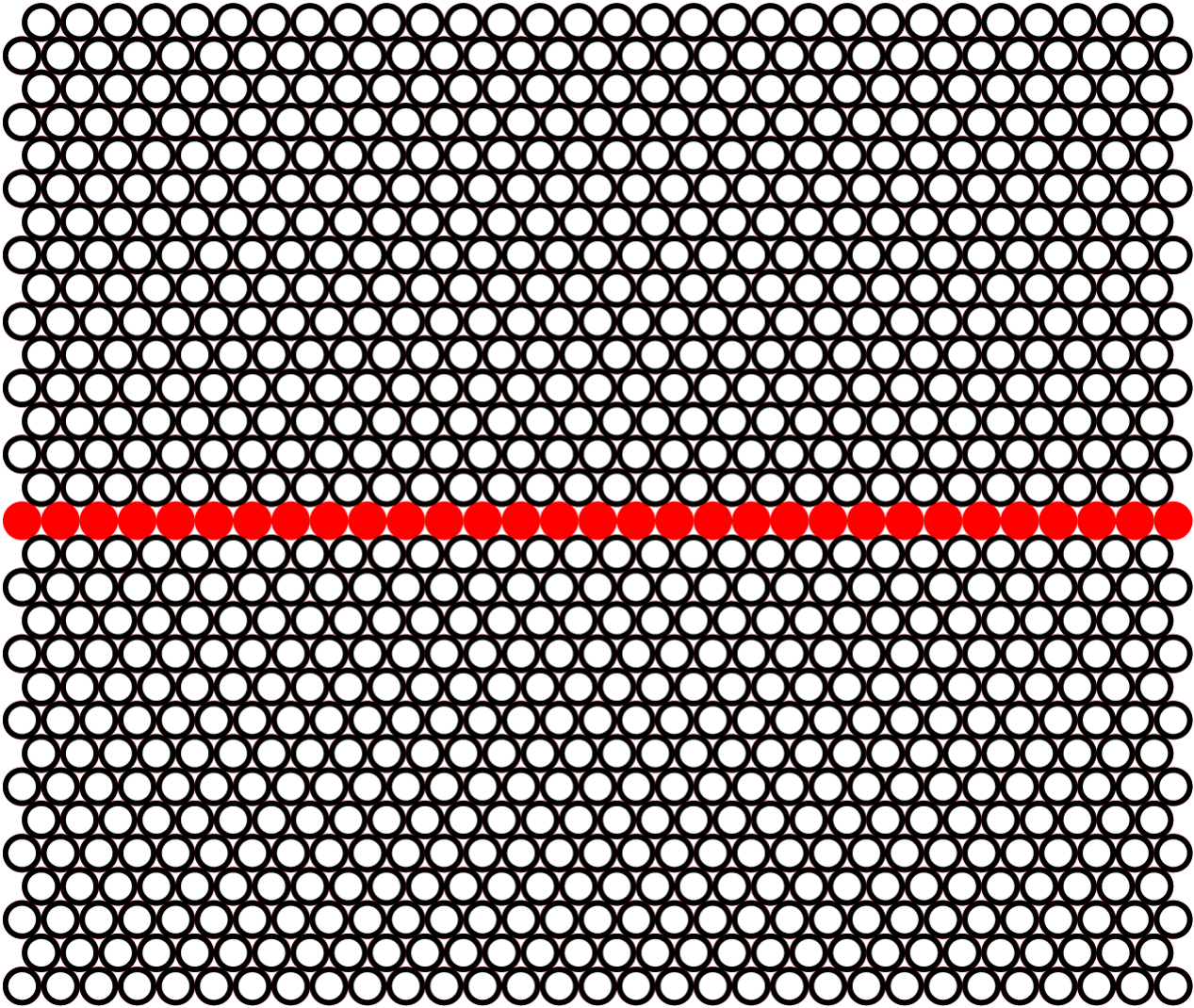}
\hspace{0.25cm}
(5)
\includegraphics[clip=true,width=4.5cm]{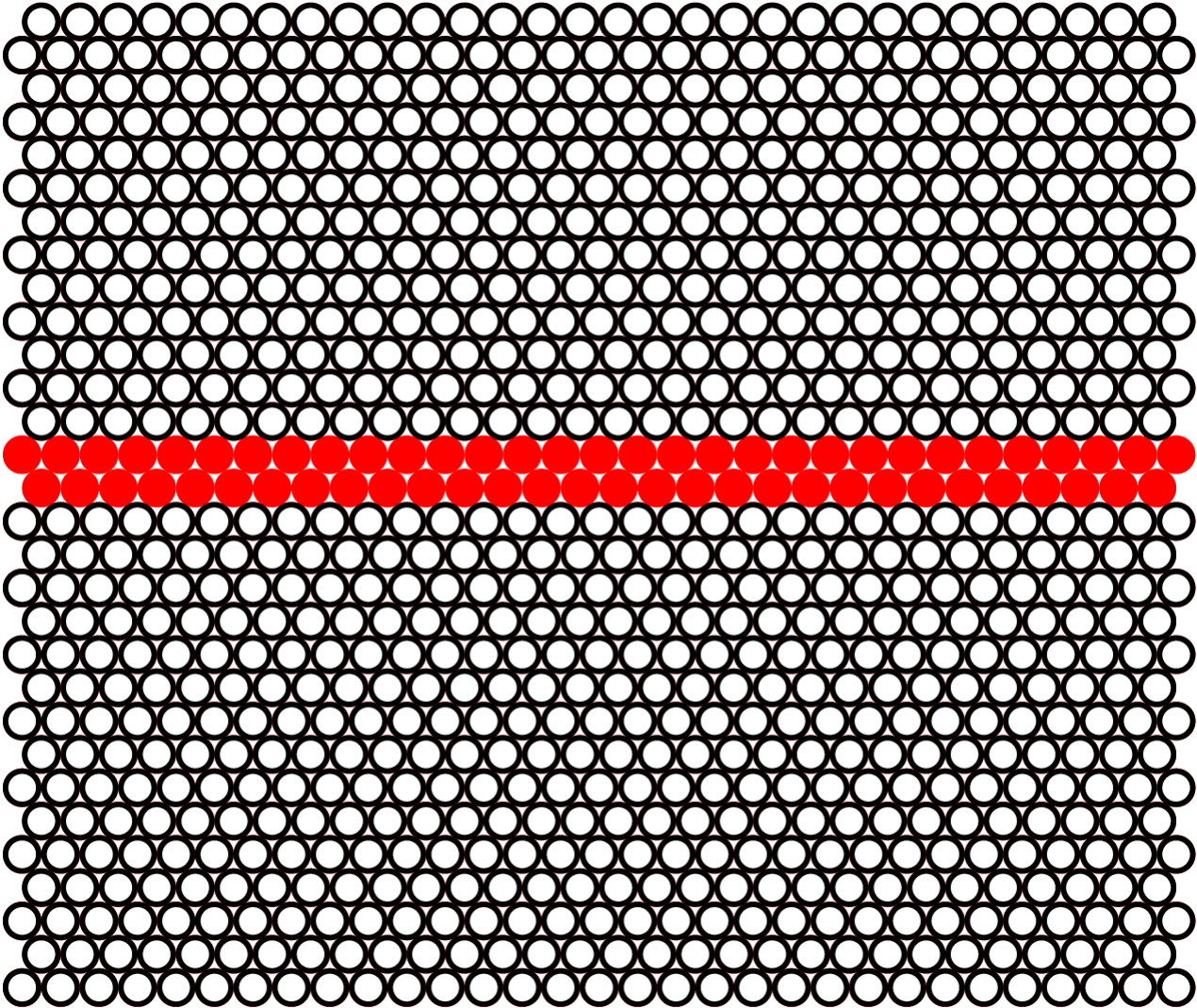}
\hspace{0.25cm}
(6)
\includegraphics[clip=true,width=4.5cm]{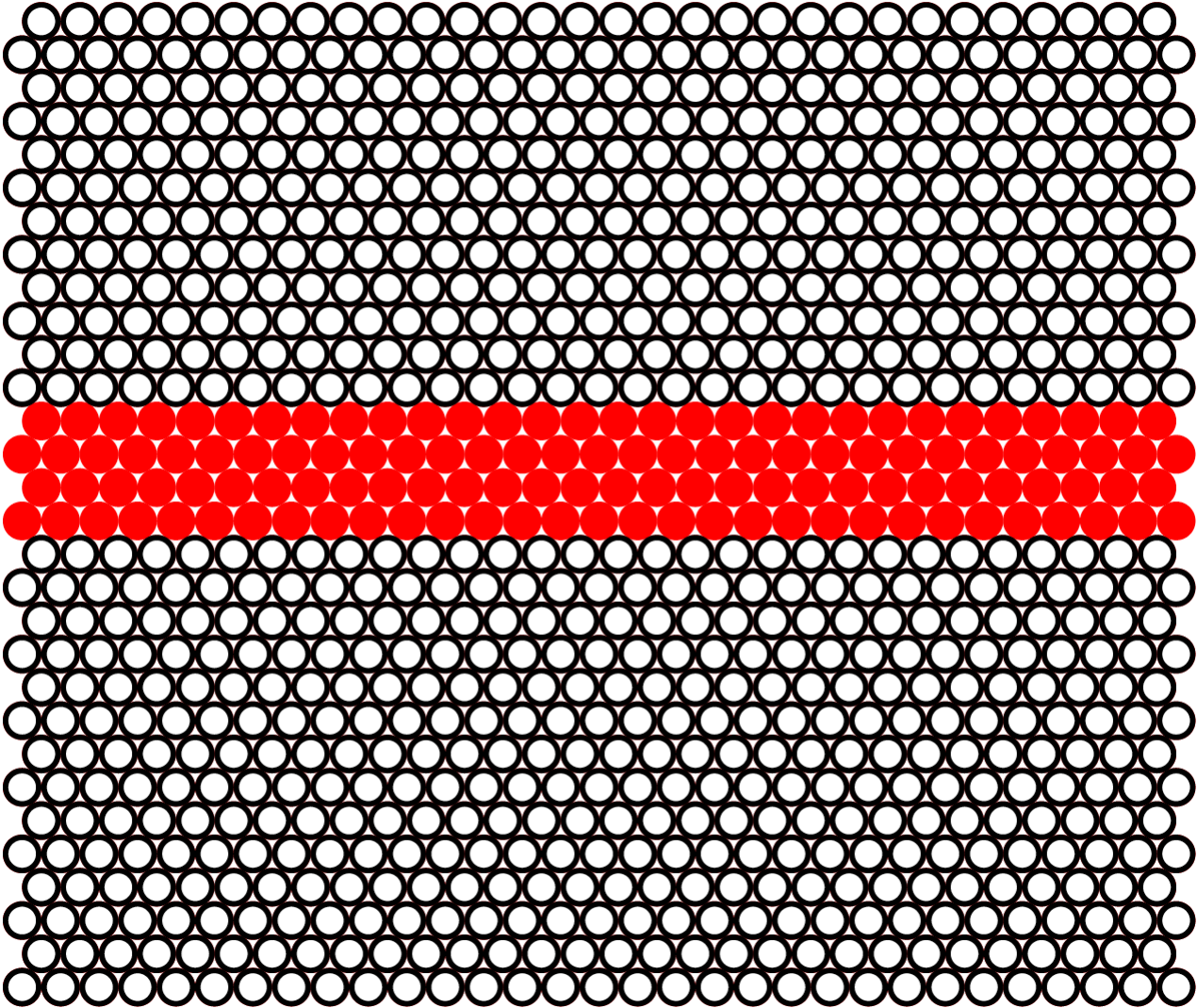}
\end{center}
\caption{Snapshots of a sub-region within a full layer composed of a triangular lattice with various defect types. The full layer measures 200x173.2, while the sub-region displayed here has dimensions 30x26 in units of $a$. The defect concentration is uniformly set at $2\%$ across the full lattice, for all defect types. Defects are shown as solid red circles.
The defect types include: (1) randomly distributed point defects, (2) randomly distributed 4-point holes, (3) randomly distributed 8-point holes, (4) 4 single-particle-width stripes spaced at $173.2/4$, (5) 2 double-particle-width stripes spaced at $173.2/2$, (6) 1 quadruple-particle-width stripe, with only one stripe shown in the sub-region.}   
\label{figsnapshots_defects}
\end{figure*}

\clearpage
\newpage

\end{document}